\documentclass[12pt]{article}
\usepackage{epsfig}
\usepackage{amssymb}
\usepackage{amsfonts}
\usepackage{subfigure}
\def\beq{\begin{equation}}
\def\eeq{\end{equation}}
\def\bea{\begin{eqnarray}}
\def\eea{\end{eqnarray}}
\def\ksl{\hbox{\hbox{${k}$}}\kern-1.9mm{\hbox{${/}$}}}

\newcommand{\nn}{\nonumber}
\newcommand{\text}{\rm}

\newcommand{\psl}{p \! \! \!  /}
\newcommand{\tRE}{\widetilde{\rm Re}}

\def\lsim{\raise0.3ex\hbox{$\;<$\kern-0.75em\raise-1.1ex\hbox{$\sim\;$}}} 

\def\gsim{\raise0.3ex\hbox{$\;>$\kern-0.75em\raise-1.1ex\hbox{$\sim\;$}}}

\begin{document}

\begin{center}

{\bf \Large One loop Standard Model corrections to 
flavor diagonal  \\
\vspace{0.3cm} 
fermion-graviton vertices} 

\vspace{1.5cm}
{\bf Claudio Corian\`{o}$^{a}$, Luigi Delle Rose$^{a}$, 
Emidio Gabrielli$^{b,c}$\footnote{
On leave of absence from Dipartimento di Fisica  Universit\`a di 
Trieste, Strada Costiera 11, I-34151 Trieste \\}, and Luca Trentadue$^{d}$}
\vspace{1cm}

{\it $^{(a)}$Dipartimento di Matematica e Fisica "Ennio De Giorgi", 
Universit\`{a} del Salento and \\ INFN-Lecce, Via Arnesano, 73100 Lecce, Italy\footnote{claudio.coriano@le.infn.it, luigi.dellerose@le.infn.it, 
emidio.gabrielli@cern.ch, luca.trentadue@cern.ch}}\\

\vspace{1cm}
{\it $^{(b)}$ NICPB, R\"avala 10, Tallinn 10143, Estonia\\
$^{(c)}$ INFN, Sezione di Trieste, Via Valerio 2, I-34127 Trieste, Italy \\}
\vspace{1cm}
{\it$^{(d)}$ Dipartimento di Fisica e Scienze della Terra "Macedonio Melloni", Universit\`a di Parma and
INFN, Sezione di Milano Bicocca, Milano, Italy
\\
}
\vspace{.5cm}
\begin{abstract} We extend a previous analysis of 
flavor-changing fermion-graviton vertices, by adding the 
one-loop SM corrections to the flavor diagonal fermion-graviton interactions. 
Explicit analytical expressions taking into account fermion masses 
for the on-shell form factors are computed and presented.
The infrared safety of the fermion-graviton 
vertices against radiative corrections of soft photons and gluons is proved, 
by extending the ordinary infrared cancellation mechanism 
between real and virtual emissions to the gravity case.
These results can be easily generalized to fermion couplings with
massive gravitons, graviscalar, and dilaton fields, with
potential phenomenological implications to new physics scenarios with
low gravity scale.

\end{abstract}
\end{center}

\newpage
\section{Introduction}
The investigation of the perturbative couplings of ordinary field theories to gravitational backgrounds has received a certain attention 
in the past \cite{'tHooft:1974bx,Capper:1975ig,Berends:1974gk,Drummond:1979pp}.
While the smallness of the gravitational coupling may shed some doubts on the practical relevance of such corrections, with the advent of models on large (universal \cite{ArkaniHamed:1998rs, Antoniadis:1998ig, ArkaniHamed:1998nn} and warped \cite{Randall:1999ee,Randall:1999vf}) extra dimensions and, more in general, of models with a low gravity scale \cite{Dvali:2000hp} their case has found a new and widespread support. 
This renewed interest covers both theoretical and phenomenological aspects that
 have not been investigated in the past. They could play, for instance, a significant role in addressing issues such as the universality of the gravitational coupling to matter \cite{Degrassi:2008mw}, in connection with the Lagrangian of the Standard Model. On the more formal side, as pointed out in some studies \cite{Giannotti:2008cv,Mottola:2006ew,Armillis:2009im,Armillis:2009pq,Armillis:2010qk}, the structure of the effective action,
accounting for anomaly mediation between the Standard Model and gravity, 
shows, in its perturbative expansion, 
the appearance of new effective scalar degrees of freedom of dilaton type. These aspects went unnoticed before and, 
for instance, could be significant in a cosmological context. 
For this reason, we believe, they require further consideration.

All these perturbative analyzes are usually performed at the leading order in the gravitational coupling. This is due to the rather involved 
expression of the operator responsible for such a coupling at classical and hence at quantum level: the energy-momentum tensor (EMT) of the Standard Model. Its expression in the electroweak (EW) theory is, indeed, very lengthy, and the classification of the several amplitudes (and form factors) in which it appears - at leading order in the EW expansion - requires a considerable effort.
 
 The previous vertices discussed in the literature have been
the graviton-fermion-antifermion vertex  ($T f \bar{f}'$) and the graviton-gauge-gauge vertex ($TJJ'$), with $J$ being a generic neutral current. A general discussion of all the amplitudes and the renormalization properties of these vertices has been presented in the $TJJ'$  case in \cite{Coriano:2011zk}, while the $T f \bar{f}'$ case has been analyzed, for fermions of different flavors, in \cite{Degrassi:2008mw}. This second work also contains a detailed discussion of the direct implications of the result for massive gravity, leading, 
in this respect, to interesting noteworthy conclusions.  
In particular, it was shown that 
flavor-changing interactions coupled to a gravity background are local only if the graviton is strictly massless, with a long-range contribution appearing in the Newton potential only if a graviton has a small mass.

In the current work we are going to revive this program and extend this previous analysis of the $T f \bar{f}'$ vertex 
\cite{Degrassi:2008mw} to the flavor diagonal case, presenting the explicit results for the EW and strong corrections to the corresponding form factors. These are given in a basis which partly overlaps with the previous tensor basis of the flavor-changing case \cite{Degrassi:2008mw}, but with some new additions.

We anticipate that these results are relevant for some phenomenological 
consequences that may affect studies on these interactions. One of the
 special contexts, which we plan to address in the future, 
 is represented by the neutrino sector.
Indeed, the analysis of the effects of the one-loop
EW radiative corrections to 
the gravitational interactions of neutrinos is missing in the literature.
Although there are previous 
studies dealing with more general structures for the 
neutrino-graviton interactions beyond the tree-level approximation, see 
 \cite{Menon:2008wa} and references therein, none of these works takes 
into account the exact neutrino-graviton vertex at one-loop 
level in the SM, including the neutrino mass dependence. 
In particular, these corrections 
are expected to play some role in the neutrino physics 
in the presence of a gravitational background, mainly due to the new 
parity-violating structures induced by EW corrections.
In this respect, we are planning to analyze in a forthcoming paper, the impact of the one-loop EW radiative 
corrections on the gravitational interactions of neutrinos, including the
effects of the EW contributions to the off-diagonal flavor transitions 
\cite{Degrassi:2008mw,new}. The inclusion of these corrections in the fermion-graviton vertex
might also help in clarifying
whether Majorana or Dirac fermions behave differently with respect to the gravitational interactions, a much debated topic in the litarature \cite{Menon:2008wa}.
Moreover, it is also left open the possibility of extending our analysis to more general gravitational backgrounds, with the inclusion of a dilaton field.

Our work is organized as follows. We briefly discuss the structure of the embedding of the Standard Model Lagrangian in a curved space-time, assuming as a background metric the usual 4-dimensional one. The discussion is rather general, and remains valid also for more general cases, which include dilaton backgrounds obtained from the compactification of higher dimensional metrics. Then we turn in section \ref{Sec.PertExp} to illustrate the structure of the perturbative expansion, organized in terms of the various contributions to the EMT of the Standard Model. These are separated with respect to the particles running in the loops, which are the photon, the W and Z gauge bosons and the Higgs field.  We conclude this section with a classification of the relevant form factors. 

In section \ref{Sec.WardId} we briefly discuss the derivation of an important Ward identity involving the effective action which is crucial to test the correctness of 
our results and to secure their consistency. Section \ref{Sec.Ren} addresses the issue of the renormalization of the theory, 
which complements the analysis of \cite{Coriano:2012nm}. We recall that no new counterterms are needed - except for those of the Standard Model Lagrangian - to carry the perturbative expansion of EMT insertions on correlators of the Standard Model, under a certain condition. This condition requires that the non-minimal coupling ($\chi$) of the Higgs field be fixed at the conformal value $1/6$. We then proceed in section \ref{Sec.FormFac} with a description of the expression of the form 
factors for each separate gauge/Higgs contribution in the loop corrections. 

In section \ref{Sec.Infrared} we give a simple proof - at leading order in the gauge couplings - of the infrared finiteness of these loop corrections, once they are combined with the corresponding real emissions of massless gauge bosons, integrated over phase space. The proof is a generalization of the ordinary cancellation between real and virtual emissions, in inclusive cross sections, to the gravity case. Finally, in section \ref{Sec.Conclusions} we give our conclusions.

\section{Theoretical framework}
\label{Sec.TheorFram}
 We recall that the dynamics of the Standard Model plus gravity is described by the Lagrangian  
\beq S = S_G + S_{SM} + S_{I}= -\frac{1}{\kappa^2}\int d^4 x \sqrt{-{g}}\, R+ \int d^4 x
\sqrt{-{g}}\mathcal{L}_{SM} + \chi \int d^4 x \sqrt{-{g}}\, R \, H^\dag H      \, ,
\label{thelagrangian}
\eeq
This includes, beside the Einstein term $\mathcal{S}_G$ where $R$ is the Ricci scalar, the $\mathcal{S}_{SM}$ action and a $\mathcal{S}_I$ term involving the Higgs doublet $H$. The latter is responsible for generating a symmetric and traceless energy-momentum tensor 
\cite{Callan:1970ze}, also called  
"term of improvement". $\mathcal{S}_{SM}$, instead, is obtained by extending the ordinary Lagrangian of the Standard Model to a curved metric background. 

Notice that $\mathcal{S}_I$ vanishes in the flat space time limit, due to the vanishing of the Ricci scalar in the same limit, but the EMT 
which is derived from it $(T_{I\,\mu\nu}$) is non-vanishing. The term $\chi$ is a parameter which remains arbitrary and that at a special value $(\chi\equiv\chi_c=1/6)$, as we have mentioned above, guarantees the renormalizability of the model at leading order in the expansion in $\mathcal{\kappa}$.  The value $\chi=\chi_c$ is usually termed "conformal coupling", in close analogy to the value necessary for the Lagrangian of a gravity-coupled scalar to exhibit conformal symmetry. Notice that the Standard Model Lagrangian is not scale invariant, due to the quadratic 
term of the Higgs field in the scalar potential, but would be such if this term were omitted. 
We will work in the almost flat space time limit, in which deviations from the flat metric  
$\eta_{\mu\nu}=(+,-,-,-)$ are parametrized in terms of the gravitational coupling $\kappa$, with $\kappa^2= 16 \pi G_N$ and with $G_N$ being the gravitational Newton's constant. At this order the metric is given as $g_{\mu\nu}=\eta_{\mu\nu} + \kappa h_{\mu\nu}$, with $h_{\mu\nu}$ denoting the fluctuations of the external graviton.  

The interaction between the gravitational field and matter, at this order, is mediated by diagrams containing a single power of the energy momentum tensor (EMT) $T^{\mu\nu}$ and multiple fields of the Standard Model. The tree level coupling is summarized by the action 
\beq
\mathcal{S}_{int}=-\frac{\kappa}{2}\int d^4 x \, T_{\mu\nu} h^{\mu\nu} \,,
\label{inter}
\eeq
where $T_{\mu \nu}$ denotes the symmetric and covariantly conserved EMT of the Standard Model Lagrangian, embedded in a curved space-time background and defined as 
\beq
T_{\mu\nu}=\frac{2}{\sqrt{-g}}\frac{\delta \left(S_{SM}+S_{I}\right)}{\delta g^{\mu\nu}} \bigg|_{g=\eta} \,.
\eeq 
The complete EMT of the Standard Model, including ghost and gauge-fixing contributions can be found in \cite{Coriano:2011zk}. 

The fermionic part of the EMT is obtained using the vielbein formalism. Indeed the fermions are coupled to gravity by using the spin connection $\Omega$ induced by the curved metric $g_{\mu\nu}$.
This allows the introduction of a derivative $\mathcal D_{\mu}$ which is covariant both under gauge and diffeomorphism transformations.
Generically, the Lagrangian for a fermion $(f)$ takes the form:
\bea
\mathcal L_{f} =  \frac{i}{2} \bar \psi \gamma^\mu (\mathcal D_{\mu} \psi) - \frac{i}{2} (\mathcal D_{\mu} \bar \psi) \gamma^{\mu} \psi - m \, \bar \psi \psi  \,,
\eea
where the covariant derivative is defined as $\mathcal D_{\mu} = \partial_{\mu} + A_{\mu} + \Omega_{\mu}$, with $A_{\mu}$ denoting the gauge field.
The spin connection takes the form
\bea
\Omega_{\mu} = \frac{1}{2} \sigma^{a b} V_a^{\nu} V_{b \nu ; \mu}
\eea
where $V$ is the vielbein, the semicolon denotes the gravitationally covariant derivative and $\sigma^{ab}$ are the generators of the Lorentz group in the spinorial representation. The latin indices are Lorentz indices of a local free-falling frame, as in Cartan's formulation. 

In establishing the Feynman rules for the perturbative expansion around an almost flat background, it is convenient to work in a gauge in which the bilinear mixing terms between gauge bosons and their longitudinal Goldstone fields, in the broken EW phase, are removed. This induces some modifications respect to the $R_{\xi}$ gauge, usually chosen in the computations of EW corrections in the flat space-time. In fact, the gauge-fixing Lagrangian in a curved gravitational background acquires a new contribution not present in the case of flat space. This is due to the promotion of the ordinary flat derivative to a covariant one \cite{Coriano:2011zk}, which requires the addition of the Christoffel connection  $\Gamma_{\mu\nu}^{\alpha}$
\bea
\partial_\mu A_\nu  \rightarrow \mathcal D_\mu = \partial_\mu A_\nu -  \Gamma_{\mu\nu}^{\alpha}A_\alpha \,.
\eea
This is the only term which needs to be varied in the gauge field Lagrangian - together with the determinant of the metric $\sqrt{-g}$ -
respect to the background field $g_{\mu\nu}$. The field strengths, for instance, are not affected by the Christoffel connection because of their antisymmetry on the Lorentz indices.  The gauge-fixing Lagrangian is then given by
\bea
\mathcal L_{g.fix} = - \frac{1}{2 \xi} \sum_{a = 0}^3 (\mathcal F^a )^2
\eea
where the gauge-fixing functions are defined as
\bea
\mathcal F^0 &=& g^{\mu\nu} \left( \partial_{\mu} B_{\nu}  - \Gamma_{\mu\nu}^{\alpha}B_\alpha \right)  + \frac{ i \xi g'}{2} \left( H^\dag \langle H \rangle -  \langle H^\dag \rangle H \right) \,, \nn \\
\mathcal F^i &=& g^{\mu\nu} \left( \partial_{\mu} W^i_{\nu}  - \Gamma_{\mu\nu}^{\alpha}W^i_\alpha \right)  + \frac{ i \xi g}{2} \left( H^\dag \sigma^i \langle H \rangle -  \langle H^\dag \rangle \sigma^i H \right) \qquad i = 1,2,3 \,.
\eea
In the previous equations $g$, $g'$, $W^i_\mu$ and $B_\mu$ are the coupling constants and the fields of the $SU(2)_W$ and $U(1)_Y$ gauge groups respectively, while $\sigma^i$ are the Pauli matrices and $\langle H \rangle$ is the vacuum expectation value of the Higgs doublet.

As we have already discussed, the non-minimal coupling of the Higgs scalar, accounted for by the Lagrangian $\mathcal{S}_I$, generates an extra contributions to the EMT. For this purpose we recall that in the broken EW phase, the ordinary parametrizations of the Higgs field  
\beq
H = \left(\begin{array}{c} -i \phi^{+} \\ \frac{1}{\sqrt{2}}(v + h + i \phi) \end{array}\right)
\eeq
and of its conjugate $H^\dagger$,
are expressed in terms of $h$, $\phi$ and $\phi^{\pm}$, which denote the physical Higgs and the Goldstone bosons of the $Z$ and $W'$ s respectively. $v$ denotes, as usual, the Higgs vacuum expectation value. This expansion 
generates a non-vanishing EMT, induced by $S_I$, given by
\bea
\label{Timpr}
T^{\mu\nu}_I = - 2 \chi (\partial^\mu \partial^\nu - \eta^{\mu \nu} \Box) H^\dag H = - 2 \chi (\partial^\mu \partial^\nu - \eta^{\mu \nu} \Box) \left( \frac{h^2}{2} + \frac{\phi^2}{2} + \phi^+ \phi^- + v \, h\right) \,.
\eea
As we have already mentioned, the most important aspect of the $\chi=1/6$ case is the renormalizability of Green's functions with an insertion of EMT and scalar fields on the external lines. These are found to be ultraviolet finite only if $T^{\mu\nu}_I$ is included \cite{Callan:1970ze,Freedman:1974gs,Coriano:2011zk}. Our results, however, are presented for an arbitrary $\chi$.

\section{The perturbative expansion}
\label{Sec.PertExp}
We will be dealing with the $Tf \bar{f}$ (diagonal) fermion case. We introduce the following notation
\bea
\hat T^{\mu\nu} = i \langle p_2 | T^{\mu\nu}(0) | p_1 \rangle
\eea
to denote the general structure of the transition amplitude where the initial and final fermion states are defined with momenta $p_1$ and $p_2$ respectively. The external fermions are taken on mass shell and of equal mass $p_1^2 = p_2^2 = m^2$. We will be using the two linear combinations of momenta  $p = p_1 + p_2$ and $q = p_1 - p_2$  throughout the paper in order to simplify the structure of the final result.

The tree-level Feynman rules needed for the computation of the $\hat T^{\mu\nu} $ vertex are listed in Appendix \ref{feynrules}, and its expression at Born level is given by
\bea
\label{treeT}
\hat T^{\mu\nu}_0 = \frac{i}{4} \bar u(p_2) \bigg\{ \gamma^{\mu} p^{\nu} + \gamma^{\nu} p^{\mu} \bigg\} u(p_1) \,.
\eea
Our analysis will be performed at leading order in the weak coupling expansion, and we will define a suitable set of independent tensor amplitudes 
(and corresponding form factors) to parameterize the result. \\
The external fermions can be leptons or quarks. In the latter case, since the EMT does not carry any non-abelian charge, the color matrix is diagonal and for notational simplicity, will not be included.

\begin{figure}[t]
\centering
\subfigure[]{\includegraphics[scale=0.8]{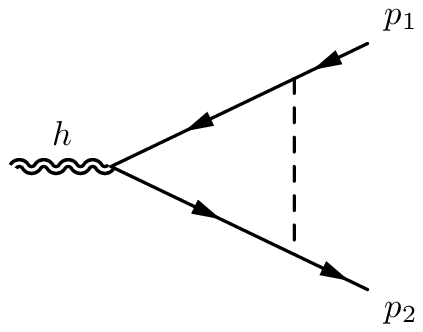}} \hspace{.5cm}
\subfigure[]{\includegraphics[scale=0.8]{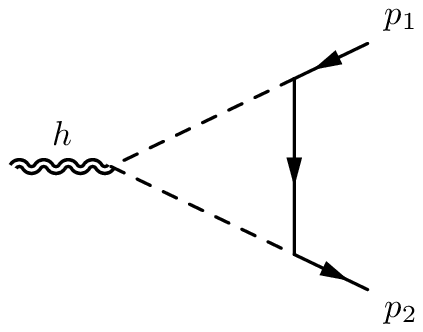}} \hspace{.5cm}
\subfigure[]{\includegraphics[scale=0.8]{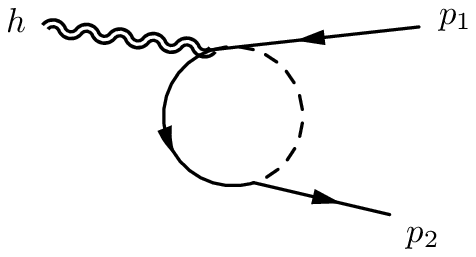}} \hspace{.5cm}
\subfigure[]{\includegraphics[scale=0.8]{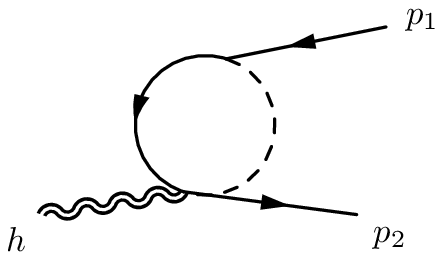}}
\caption{The one-loop Feynman diagrams of the graviton fermion vertex. The dashed lines can be gluons, photons, Higgs, $Z$ and $W$ bosons or their unphysical longitudinal parts. The internal fermion line can be of the same flavor of the external fermions if a neutral boson is exchanged in the loop, otherwise, for charged $W$ corrections, it can have different flavor because of the CKM matrix. \label{diagrams}}
\end{figure}

We decomposed the full matrix element into six different contribution characterized by the SM sectors running in the loop diagrams
\bea
\label{hatT}
\hat T^{\mu\nu} = \hat T^{\mu\nu}_{g} + \hat T^{\mu\nu}_{\gamma} + \hat T^{\mu\nu}_{h} + \hat T^{\mu\nu}_{Z} + \hat T^{\mu\nu}_{W} + \hat T^{\mu\nu}_{CT} 
\eea
where the subscripts stand respectively for the gluon, the photon, the Higgs, the $Z$ and the $W$ bosons and the counterterm contribution. Concerning the last term we postpone a complete discussion of the vertex renormalization to 
a follow-up section. \\
As we have already mentioned, we work in the $R_\xi$ gauge, where the sector of each massive gauge boson is always accompanied by the corresponding unphysical longitudinal part. This implies that the diagrammatic expansion of $\hat T^{\mu\nu}_{Z}$ and $\hat T^{\mu\nu}_{W}$ is characterized by a set of gauge boson running in the loops with duplicates obtained by replacing the massive gauge fields with their corresponding Goldstones. \\
The decomposition in Eq.(\ref{hatT}) fully accounts for the SM one-loop corrections to the flavor diagonal EMT matrix element with two external fermions. \\
The various diagrammatic contributions appearing in the perturbative expansion are shown in Fig.\ref{diagrams}. Two of them are characterized by a typical triangle topology, while the others denote terms where the insertion of the EMT and the fermion field occur on the same point. The computation of these diagrams is rather involved and has been performed in dimensional regularization using the on-shell renormalization scheme. We have used the standard reduction of tensor integrals to a basis of scalar integrals and we have checked explicitly the Ward identity coming from the conservation of the EMT, which are crucial to secure the correctness of the computation.  \\
\subsection{Tensor decompositions and form factors}
Now we illustrate in more detail the organization of our results. By using symmetry arguments and exploiting some consequences of the Ward identities, we have determined a suitable tensor basis on which our results are expanded. For massless vector bosons (gluons and photons), and for the Higgs field, because of the parity-conserving nature of their interactions, we have decomposed the matrix elements onto a basis of four tensor structures $O^{\mu\nu}_{V  k}$ with four form factors $f_k$ as

\bea
\hat T^{\mu\nu}_{g} &=& i \frac{\alpha_s }{4 \pi} C_2(N)  \sum_{k=1}^4 f_k(q^2) \, \bar u(p_2) \, O^{\mu\nu}_{V  k} \, u(p_1) \,, \\
\hat T^{\mu\nu}_{\gamma} &=& i \frac{\alpha }{4 \pi} Q^2  \sum_{k=1}^4 f_k(q^2) \, \bar u(p_2) \, O^{\mu\nu}_{V k} \, u(p_1) \,, \\
\hat T^{\mu\nu}_h &=& i  \frac{G_F}{16 \pi^2 \sqrt{2}}  \, m^2  \sum_{k=1}^4 f^h_k(q^2) \, \bar u(p_2) \, O^{\mu\nu}_{V  k} \, u(p_1) \,,
\eea
with the tensor basis defined as
\bea
\label{vectorbasis}
O^{\mu\nu}_{V  1} &=& \gamma^\mu \, p^\nu + \gamma^\nu \, p^\mu \,, \nn \\
O^{\mu\nu}_{V 2} &=& m \, \eta^{\mu\nu} \,, \nn \\
O^{\mu\nu}_{V  3} &=& m \, p^\mu \, p^\nu \,, \nn \\
O^{\mu\nu}_{V 4} &=& m \, q^\mu \, q^\nu \,.
\eea 
The form factors for the gluon and the photon contributions are identical, the only difference relies in the coupling constant and in the charge of the external fermions. The coefficient $C_2(N)$ is the quadratic Casimir in the $N$-dimensional fundamental representation, with $C_2 = 4/3$ for quarks and zero for leptons. $Q$ denotes the electromagnetic charge and $G_F$ the Fermi constant. Moreover, being the fermion-Higgs coupling proportional to the fermion mass, we have factorized $m^2$ in front of the Higgs form factors. Note that $O^{\mu\nu}_{V  2-4}$ are linearly mass suppressed, so that only $O^{\mu\nu}_{V  1}$ survives in the limit of massless external fermions. \\
Coming to the weak sector of our corrections, because of the chiral nature of the $Z$ and $W$ interactions, we have to decompose the matrix elements into a more complicated tensor basis of six elements as
\bea
\hat T^{\mu\nu}_Z &=& i \, \frac{G_F}{16 \pi^2 \sqrt{2}}   \sum_{k=1}^6  f^{Z}_k(q^2) \, \bar u(p_2) \, O^{\mu\nu}_{C  k} \, u(p_1) \,, \\ 
\hat T^{\mu\nu}_W &=& i \, \frac{G_F}{16 \pi^2 \sqrt{2}}  \sum_{k=1}^6  f^{W}_k(q^2) \, \bar u(p_2) \, O^{\mu\nu}_{C  k} \, u(p_1) \,, 
\eea
where we have defined
\bea
\label{chiralbasis}
O^{\mu\nu}_{C  1} &=& \left( \gamma^\mu \, p^\nu + \gamma^\nu \, p^\mu \right) P_L \,, \nn \\
O^{\mu\nu}_{C  2} &=& \left( \gamma^\mu \, p^\nu + \gamma^\nu \, p^\mu \right) P_R \,, \nn \\
O^{\mu\nu}_{C  3} &=& m \, \eta^{\mu\nu} \,, \nn \\
O^{\mu\nu}_{C  4} &=& m \, p^\mu \, p^\nu  \,, \nn \\
O^{\mu\nu}_{C  5} &=& m \, q^\mu \, q^\nu  \,, \nn \\
O^{\mu\nu}_{C  6} &=& m \, \left( p^\mu \, q^\nu + q^\mu \, p^\nu \right) \gamma^5 \,.
\eea
The most general rank-2 tensor basis that can be built with a metric tensor, two momenta ($p$ and $q$) and matrices $\gamma^\mu, \gamma^5$ has been given in \cite{Degrassi:2008mw}. The basis given in Eq. (\ref{chiralbasis}), compared to the flavor-changing case, is more compact.  We have imposed the  symmetry constraints on the external fermion states (of equal mass and flavor) and the conservation of the EMT, discussed in section \ref{Sec.WardId}.
For the form factors appearing in $\hat T^{\mu\nu}_W$ we introduce the notation  
\bea
\label{Ftof}
f^{W}_k(q^2) =  \sum_f V_{if}^* V_{fi} \, F^{W}_k(q^2, x_f) 
\eea
where $V_{i f}$ is the CKM mixing matrix, with the indices $i$ and $f$ 
corresponding to the flavor of the external and internal-loop fermions
respectively, and $x_f=m_f^2/m_W^2$, where $m_f$ and $m_W$ stand for
the masses of the fermion $f$ and $W$ respectively. \\
We have extracted a single mass suppression factor coming from the contribution of the $O^{\mu\nu}_{C 3-6}$ operators. In the $\hat T^{\mu\nu}_Z$ matrix element, the leading terms in the small external fermion mass are given by the first two form factors. The situation is different for the $W$ case, in which only the first form factor is the leading term, being $f_2$ suppressed as $m^2$. We have decided not to factorize the $m^2$ term in order to make the notations uniform with the $Z$ case. \\
We remark that the expressions of the form factors is exact, having kept in the result the complete dependence from all the kinematic invariants and from the external and internal masses.

\section{The Ward identity from the conservation of the EMT}
\label{Sec.WardId}
In this section we simply quote the consequences of the conservation of the energy-momentum tensor that are contained in the Ward identities satisfied by the matrix elements defined above. As widely explained in \cite{Coriano:2011zk}, we can derive a master equation for the effective action $\Gamma$, the generating functional of all the 1-particle irreducible (1PI) graphs. We give more details of the derivation in Appendix \ref{WIS}. The Ward identities for the various correlators are then obtained via functional differentiation. We consider a generic theory with a scalar $\phi$, a gauge field $A_{\alpha}$ and a fermion $\psi$. As explained in the Appendix \ref{WIS}, $\phi^c$ can be taken to represents all the scalar and ghosts contributions, while $A_{\alpha}$ is a short-hand notation to indicate all the contributions from the gauge bosons. \\
Imposing the invariance of the generating functional under a diffeomorphism transformation of $\Gamma$ we have  
\bea
\label{masterWI}
\partial_{\mu} \frac{\delta \Gamma}{h_{\mu\nu}} &=& - \frac{\kappa}{2} \bigg\{  - \frac{\delta \Gamma}{\delta \phi_c} \partial^{\nu} \phi_c - \partial^{\nu} \phi_c^\dag \frac{\delta \Gamma}{\delta \phi_c^\dag} - \frac{\delta \Gamma}{\delta A_{c\, \alpha}} \partial^{\nu} A_{c \, \alpha}  + \partial_{\alpha} \left( \frac{\delta \Gamma}{\delta A_{c \, \alpha }}  A_c^{\nu}  \right) \nn \\
&-& \partial^{\nu} \bar \psi_c  \frac{\delta \Gamma}{\delta \bar \psi_c}  -   \frac{\delta \Gamma}{\delta \psi_c} \partial^{\nu} \psi_c  + \frac{1}{2} \partial_{\alpha} \left( \frac{\delta \Gamma}{\delta \psi_c}  \sigma^{\alpha \nu} \psi_c -  \bar \psi_c \sigma^{\alpha \nu} \frac{\delta \Gamma}{\delta \bar \psi_c}  \right) \bigg\} \,,
\label{classical}
\eea
where $\sigma^{\alpha \beta} = [\gamma^{\alpha}, \gamma^{\beta}]/4$ and the subscript $c$ identifies the classical fields. This equation can be straightforwardly generalized to the entire spectrum of the SM. \\
By a functional differentiation of Eq.(\ref{masterWI}) with respect to the fermion fields and after a Fourier transform to momentum space we obtain
\bea
\label{WI}
q_{\mu} \, \hat T^{\mu\nu}& =& \bar u(p_2) \bigg\{ p_2^{\nu} \,  \Gamma_{\bar f f}(p_1) -  p_1^{\nu} \, \Gamma_{\bar f f}(p_2)
 + \frac{q_\mu}{2} \left( \Gamma_{\bar f f}(p_2) \, \sigma^{\mu\nu} - \sigma^{\mu\nu} \, \Gamma_{\bar f f}(p_1) \right) \bigg\} u(p_1) \,,
\eea
where $ \Gamma_{\bar f f}(p)$ is the fermion two-point function, diagonal in flavor space, given explicitly in Appendix \ref{selfenergies}. The perturbative test of this equation is of great importance for testing the correctness of our results. 

At this point we would like to stress that, as a strong test of our results, 
we have explicitly 
computed all the form factors entering in the matrix elements of
$\hat{T}_{g,h,\gamma,W,Z}^{\mu\nu}$ and checked that they 
satisfy the Ward identity in Eq.(\ref{WI}). 
However, as a consequence of  Eq.(\ref{WI}), not all the form 
factors are independent quantities.  Therefore, for practical text
purposes,  we will present only the analytical results 
for the relevant independent subset of form factors,
thus reducing the number of contributions to the $T f\bar{f}$ vertex.
The other form factors can be derived quite straightforwardly by using the
Ward identity given above, which determines a set of relations among them that will be presented in the next sections.

\section{Renormalization}
\label{Sec.Ren}

The counterterms needed for the renormalization of the vertex can be obtained by promoting the counterterm Lagrangian to the curved background. The counterterm Feynman rules for the matrix element with the insertion of the EMT are easily extracted in the usual way and in our case, for a chiral fermion, we have
\bea
\label{TCT}
\hat T^{\mu \nu}_{CT} = i \langle p_2 | T^{\mu\nu}_{CT} (0) | p_1 \rangle =  \frac{i}{4} \bar u(p_2) \bigg\{ \delta Z_L \, O^{\mu\nu}_{C 1}   +  \delta Z_R \, O^{\mu\nu}_{C  2}  + 4 \frac{\delta m}{m} O^{\mu\nu}_{C  3}  \bigg\} u(p_1) \,,
\eea
with $\delta Z_L$, $\delta Z_R$ and $\delta m$ being the fermion wave function and the mass renormalization 
constants respectively, while $O^{\mu\nu}_{C 1-3}$ are defined in Eq.(\ref{chiralbasis}). \\
For vector-type interactions, in the gluon, photon and Higgs sector, $\delta Z_L = \delta Z_R$ and the expansion of the $\hat T^{\mu \nu}_{CT}$ matrix element naturally collapses to only two operators, $O^{\mu\nu}_{V  1} = O^{\mu\nu}_{C  1} + O^{\mu\nu}_{C  2}$ and $O^{\mu\nu}_{V  2} = O^{\mu\nu}_{C  3}$ of Eq.(\ref{vectorbasis}). \\
We have checked that the renormalization of the parameters of the SM Lagrangian is indeed sufficient to cancel all the singularities of the $\hat T^{\mu\nu}$ matrix element, as expected. As it can be easily seen from Eq.(\ref{TCT}), the form factors involved in the subtraction of infinities are just the first two for the gluon, the photon and the Higgs, and the first three for the massive gauge bosons. This is in agreement with simple power counting arguments. \\  
We have used the on-shell scheme, where the renormalization conditions are fixed in terms of the physical parameters of the theory to all orders in the perturbative expansion in the EW coupling constants. These are the masses of physical particles, the electric charge and the CKM mixing matrix. The renormalization conditions on the fields, which allow the extraction of the wave function renormalization constants, are satisfied by requiring a unitary residue of the full 2-point functions on the physical particle poles. For the fermion renormalization constants we obtain the following explicit expressions 
\bea
\delta Z_L &=&  - \tRE \, \Sigma^L (m^2) - m^2 \frac{\partial}{\partial p^2 } \tRE \bigg[ \Sigma^L(p^2) +  \Sigma^R(p^2) +  2 \Sigma^S(p^2) \bigg]_{p^2 = m^2}  \,, \\
\delta Z_R &=&  - \tRE \, \Sigma^R (m^2) - m^2 \frac{\partial}{\partial p^2 } \tRE \bigg[ \Sigma^L(p^2) +  \Sigma^R(p^2) +  2 \Sigma^S(p^2) \bigg]_{p^2 = m^2}  \,, \\
\delta m &=& \frac{m}{2} \tRE \bigg[ \Sigma^L(m^2) +  \Sigma^R(m^2) +  2 \Sigma^S(m^2)  \bigg] \,,
\eea
where the $\Sigma^{L, R, S}$ functions are the fermion self-energies defined in Appendix \ref{selfenergies}. The symbol $\tRE$ gives the real part of the scalar integrals appearing in the self-energies but it has no effect on the CKM matrix elements. If the mixing matrix is real $\tRE$ can be replaced with $\rm Re$.

\section{Form factors for the $\hat T^{\mu\nu}$ matrix element}
\label{Sec.FormFac}

\subsection{The massless gauge boson contribution}
We give the four form factors for the massless gauge boson cases, namely the gluon and the photon contributions. They depend on the kinematic invariant $q^2$, the square of the momenta of the graviton line, and from the dimensionless ratio $y = m^2 / q^2$. The form factors are expressed as a combination of one-, two- and three-point scalar integrals, which have been defined in Appendix. \ref{scalarint}, and are given by
\small
\bea
f_1(q^2) &=&
-\frac{4 y (2 y+1)}{3 (1- 4 y)^2}
+\frac{(8 y (7 y-4)+3) }{3 q^2 (1-4 y)^2 y} \mathcal A_0 \left(m^2\right)
-\frac{2 (y-1) }{3 (1-4 y)^2} \mathcal B_0 \left(q^2,0,0\right)      \nn \\
&&  +\frac{(17-44 y) }{48 y-12} \mathcal B_0 \left(q^2 ,m^2,m^2\right)
+\frac{1}{2} q^2 (2 y-1) \mathcal C_0 \left(0,m^2,m^2 \right) 
 +\frac{2 q^2 (1-2 y) y }{(1-4 y)^2} \mathcal C_0 \left(m^2,0,0\right) \,,
 \nn \\
f_2(q^2) &=&
\frac{4 (y (32 y-23)+3)}{3 (1-4 y)^2}
+\frac{2 \left(32 y^2-26 y+3\right) }{3 q^2 (1-4 y)^2 y} \mathcal A_0 \left( m^2 \right)
+ \frac{2 (10 y-1)}{3 (1-4 y)^2}  \mathcal B_0 \left(q^2,0,0 \right) \nn \\
&& +\frac{5}{3} \mathcal B_0 \left(q^2,m^2,m^2\right)
+\frac{8 q^2 y^2 }{(1-4 y)^2} \mathcal C_0 \left(m^2,0,0\right) \,,
\nn \\
f_3(q^2) &=&
\frac{4 y (8 y+19)-6}{3 q^2 (4 y-1)^3}
-\frac{4 (8 y-17)}{3 q^4 (4 y-1)^3}  \mathcal A_0 \left( m^2 \right)
-\frac{2 (26 y+1) }{3 q^2 (4 y-1)^3} \mathcal B_0 \left(q^2,0,0\right) \nn \\
&& +\frac{2 }{3 q^2 (4 y-1)} \mathcal B_0 \left(q^2,m^2,m^2 \right)
-\frac{8 y (y+1) }{(4 y-1)^3} \mathcal C_0 \left(m^2,0,0\right) \,,
 \nn \\
f_4(q^2) &=&
\frac{-80 y^2+68 y-9}{3 q^2 (1-4 y)^2}
+\frac{(20 y (4 y-1)+3) }{3 q^4 (1-4 y)^2 y} \mathcal A_0 \left(m^2\right)
+ \frac{(2-20 y)}{3 q^2 (1-4 y)^2}  \mathcal B_0 \left(q^2,0,0\right) \nn \\
&& -\frac{5 }{3 q^2} \mathcal B_0 \left(q^2,m^2,m^2\right)
-\frac{8 y^2 }{(1-4 y)^2} \mathcal C_0 \left(m^2,0,0\right) \,.
\eea
\normalsize
It is interesting to observe that not all the four form factors are independent because the Ward identity imposes relations among them. In fact, specializing Eq.(\ref{WI}) to the massless gauge bosons contributions, we obtain
\bea
f_2(q^2) + q^2 \, f_4(q^2) = - \frac{1}{2} \bigg[ \Sigma^L_{g / \gamma}(m^2) +  \Sigma^R_{g / \gamma}(m^2)+ 2  \Sigma^S_{g / \gamma}(m^2) \bigg],
\eea
as it can be checked from the explicit expressions given above. This relation can be used to test the correctness of our results and to reduce the number of independent form factors. We recall once more that the $\Sigma$'s denote the fermion self-energies which have been collected in Appendix \ref{selfenergies}. 

\subsection{The Higgs boson contribution}
In this section we present the results for the contribution of a virtual Higgs. As in the previous case, they are expanded in terms of scalar integrals and of the kinematic variables $q^2$, $y = m^2/q^2$ and $x_h = m^2/m_h^2$, where $m_h$ is Higgs mass. \\
As we have already mentioned, the conservation of the EMT induces a Ward identity on the correlation functions. 
This implies a relation between the form factors, which in the Higgs case becomes
\bea
f^h_2(q^2) + q^2 \, f^h_4(q^2) = - \frac{1}{2} \bigg[ \Sigma^L_{h}(m^2) + \Sigma^R_{h}(m^2) + 2 \Sigma^S_{h}(m^2) \bigg] \,.
\eea
Obviously, this equation has the same structure of the Ward identity found in the massless gauge bosons case, having expanded $\hat T^{\mu\nu}_h$ on the same tensor basis of Eq.(\ref{vectorbasis}). \\
Notice that the $f_2^h$ and $f_4^h$ form factors depend on the $\chi$ parameter of $\mathcal{S}_I$. This is expected, because the Higgs field can also couple to gravity with the EMT of improvement $T^{\mu\nu}_I$ defined in Eq.(\ref{Timpr}).  The Feynman rules for a graviton-two Higgs vertex are then modified with the inclusion of the $\chi$ dependence and affect the diagram represented in Fig.\ref{diagrams}(b), where this vertex appears. We obtain 
\small
\bea
f^h_1(q^2) &=& 
\frac{3 x_h-8 y -4 x_h y }{12 x_h(1- 4y)}
+\frac{2}{3 \, q^2 (1-4 y)} \bigg[  \mathcal A_0 \left(m_h^2 \right)   -  \mathcal A_0 \left(m^2 \right)    \bigg] \nn \\
&+&  \frac{1}{12 x_h^2 (1-4  y)^2}  \bigg[ x_h^2+8 (x_h (26 x_h+3)-3) y^2-2 (28 x_h+3) x_h y\bigg] \mathcal B_0 \left(q^2,m^2, m^2\right) \nn \\
&+& \frac{1}{6 x_h^2 (1-4  y)^2} \bigg[ x_h^2+4 (3-8 x_h) y^2+4 (2 x_h-1) x_h y\bigg]  \mathcal B_0 \left(q^2,m_h^2,m_h^2 \right) \nn \\
&+& \frac{y}{2 x_h (1-4 y)^2} \bigg[ x_h (4-32 y)+12 y+1\bigg]  \mathcal B_0 \left(m^2,m^2,m_h^2 \right)
- \frac{q^2 y}{2 x_h^3 (1-4 y)^2}  \bigg[ 4 x_h^3  \nn \\
&+& 4 (x_h (4 x_h-3) (4 x_h+1)+1) y^2 
+ (8 (1-4 x_h) x_h+3) x_h y\bigg]  \mathcal C_0 \left(m_h^2,m^2,m^2 \right) \nn \\
&+& \frac{q^2 y (x_h-2 y)}{x_h^3 (1-4 y)^2}   \left(x_h^2-4 x_h y+y\right)  \mathcal C_0 \left(m^2,m_h^2, m_h^2\right)
\,, \nn \\
f^h_2(q^2) &=&
\frac{40 x_h y-9 x_h-4 y}{3 x_h(1-4 y)}
+\frac{4}{3 q^2 (1-4 y)} \bigg[  \mathcal A_0 \left(m_h^2 \right) -  \mathcal A_0 \left(m^2 \right) \bigg] 
- \frac{2}{3 x_h^2 (1-4 y)^2} \bigg[ 2 x_h^2 (1-4 y)^2 \nn \\
&+& 9 x_h y (1-4 y)+6 y^2 \bigg]    \mathcal B_0 \left(q^2,m^2,m^2\right) 
+ \frac{1}{3 x_h^2 (1-4 y)^2} \bigg[ x_h^2+4 (3-20 x_h) y^2+8 (x_h+1) x_h y\bigg]  \nn \\
&\times& \mathcal B_0 \left(q^2,m_h^2, m_h^2\right) 
+ \frac{2}{x_h (1-4 y)^2} \bigg[ 4 (1-4 x_h) y^2+6 x_h y-x_h+y\bigg]   \mathcal B_0 \left(m^2, m^2, m_h^2 \right) \nn \\
&-& \frac{4 q^2 y (-4 x_h y+x_h+y)^2 }{x_h^3 (1-4 y)^2} \mathcal C_0 \left(m_h^2 ,m^2 ,m^2 \right) \nn \\
&-& \frac{q^2 (x_h (8 y-1)-2 y)}{x_h^3 (1-4 y)^2}  \bigg[ x_h^2 (2 y-1)+8 x_h y^2-2 y^2\bigg] \mathcal C_0 \left(m^2, m_h^2, m_h^2 \right) \nn \\
&+&  \chi \bigg\{
\frac{8}{1-4 y}  \bigg[ \mathcal B_0 (m^2 , m^2,m_h^2 ) - \mathcal B_0 (q^2,m_h^2,m_h^2 ) \bigg]
+ \frac{4 q^2 (x_h+2 y -8 x_h y ) }{x_h (1-4 y)} \mathcal C_0 (m^2 ,m_h^2, m_h^2 )
\bigg\} \,, \nn \\
f^h_3(q^2) &=&
\frac{2 (x_h (22 y-3)-10 y)}{3 q^2 x_h (1-4 y)^2} 
+ \frac{2 (3-2 y) }{3 q^4 (1-4 y)^2 y} \bigg[  \mathcal A_0 \left(m_h^2 \right) - \mathcal A_0 \left(m^2 \right) \bigg] \nn \\
&+&  \frac{5 }{3 q^2 \, x_h^2 (4 y-1)^3} \bigg[ x_h^2+4 (4 (x_h-3) x_h+3) y^2+4 (3-2 x_h) x_h y\bigg] \mathcal B_0 \left(q^2 ,m^2, m^2 \right) \nn \\
&+& \frac{1}{3 q^2 \, x_h^2 (4 y-1)^3}  \bigg[ x_h^2 (7-88 y)+8 x_h y (26 y+1)-60 y^2\bigg] \mathcal B_0 \left(q^2,m_h^2,m_h^2 \right) \nn \\
&+& \frac{2}{q^2 x_h (4 y-1)^3} (x_h (2 (13-8 y) y-3)+8 (y-2) y+1)  \mathcal B_0 \left(m^2 ,m^2, m_h^2 \right) \nn \\
&+& \frac{10 y}{x_h^3 (4 y-1)^3}  (x_h (4 y-1)-2 y) (x_h (4 y-1)-y) \mathcal C_0 \left(m_h^2 ,m^2,m^2 \right) \nn \\
&+& \frac{1 }{x_h^3 (4 y-1)^3} \bigg[ x_h^3-2 \left(8 x_h^2-26 x_h+3\right) x_h y^2-2 (5 x_h+1) x_h^2 y \nn \\
&+& 4 (4 x_h-5) (4 x_h-1) y^3\bigg] \mathcal C_0 \left( m^2,m_h^2,m_h^2 \right) \,, \nn \\
f^h_4(q^2) &=&
\frac{9 x_h+4 y -40 x_h y}{3 q^2 \, x_h (1-4  y)}
+\frac{(8 y- 3) }{3 q^4 y (1 -4 y)} \bigg[ \mathcal A_0 \left(m_h^2 \right) -  \mathcal A_0 \left(m^2 \right) \bigg] 
+  \frac{2}{3 q^2 x_h^2 (1-4  y)^2} \bigg[ 2 x_h^2 (1-4 y)^2 \nn \\
&+& 9 x_h y (1-4 y)+6 y^2\bigg]  \mathcal B_0 \left(q^2 ,m^2, m^2 \right) 
- \frac{1}{3 q^2 x_h^2 (1-4  y)^2}  \bigg[ x_h^2+4 (3-20 x_h) y^2 \nn \\
&+& 8 (x_h+1) x_h y\bigg]   \mathcal B_0 \left(q^2, m_h^2, m_h^2 \right) 
+ \frac{1}{q^2 x_h (1-4 y)^2}  \bigg[ x_h (4 (5-8 y) y-2)+2 y (4 y-5)+1\bigg] \nn \\ 
&\times& \mathcal B_0 \left(m^2 ,m^2, m_h^2 \right) 
+ \frac{4 y (x_h+y -4 x_h y )^2 }{x_h^3 (1-4 y)^2} \mathcal C_0 \left(m_h^2, m^2,m^2 \right) 
+ \frac{(x_h (8 y-1)-2 y)}{x_h^3 (1-4 y)^2}  \bigg[ x_h^2 (2 y-1) \nn \\
&+& 8 x_h y^2-2 y^2 \bigg]  \mathcal C_0 \left(m^2, m_h^2, m_h^2 \right) 
+  \chi \bigg\{
\frac{8 }{q^2( 1- 4  y)} \bigg[  \mathcal B_0 \left(q^2, m_h^2, m_h^2 \right) -  \mathcal B_0 \left(m^2, m^2, m_h^2 \right) \bigg]  \nn \\
&-& \frac{4 (x_h+2 y -8 x_h y) }{x_h (1 - 4 y)}  \mathcal C_0 \left( m^2, m_h^2, m_h^2\right)
\bigg\} \,.
\eea
\normalsize

\subsection{The $Z$ gauge boson contribution}
Coming to the form factors for the $Z$ boson contribution, which are part of $\hat T^{\mu\nu}_Z$, these are given in terms of the variables $q^2$, $y=m^2/q^2$ and $x_Z \equiv m^2/m_Z^2$, with the parameters $v$ and $a$ denoting the vector and axial-vector $Z$-fermion couplings. In particular we have
\bea
\label{Zfermconst}
v = I_3 - 2 s_W^2 Q\,, \qquad a = I_3 \,, \qquad c^2 = v^2 + a^2 \,,
\eea
where $I_3$ and $Q$ are, respectively, the third component of isospin and the electric charge of the external fermions, while $s_W$ is the sine of the weak angle. 

In this case, the structure of the Ward identity is more involved than the previous cases, being $\hat T^{\mu\nu}_Z$ expanded on a more complicated tensor basis, Eq.(\ref{chiralbasis}). We obtain two relations among the form factors that we have tested on our explicit computation, which are given by
\bea
\label{WIZ1}
f^Z_2 &=&  f^Z_1 + q^2 f^Z_6 + \frac{1}{4} \bigg[ \Sigma^R_Z(m^2) -  \Sigma^L_Z(m^2)  \bigg]  \,, \\
\label{WIZ2}
f^Z_3 &=& - q^2 f^Z_5  - \frac{1}{2} \bigg[ \Sigma^L_Z(m^2) + \Sigma^R_Z(m^2) + 2 \Sigma^S_Z(m^2) \bigg] \,.
\eea
Also in this case we have a dependence of the result on the parameter $\chi$, which appears in $f^Z_5$ and hence in $f^Z_3$. As for the Higgs field, also the gravitational coupling of the $Z$ Goldstone boson acquires a new contribution coming from the term of improvement $T_I$, shown by the Feynman diagram in Fig.\ref{diagrams}(b).

Here we present a list of the explicit expressions of $f^Z_1$, $f^Z_4$, $f^Z_5$ and $f^Z_6$ while $f^Z_2$ and $f^Z_3$ can be obtained using the Ward identity constraints of Eq.(\ref{WIZ1}) and Eq.(\ref{WIZ2}). We obtain 

\small
\bea
f^Z_1(q^2) &=&
\frac{q^2 y}{3 (4 y-1) x_Z^2}   \bigg[ x_Z \left(-4 y \left(5 a^2+5 a v+7 v^2\right)+a^2 (4 y-3) x_Z+6 (a+v)^2\right)+4 y (a+v)^2\bigg] \nn \\
&+&  \frac{4 y}{3 (1 - 4 y) x_Z} \left(2 a^2 x_Z+a^2-a v+v^2\right) \bigg[ \mathcal A_0 \left( m_Z^2 \right)  -   \mathcal A_0 \left( m^2 \right) \bigg] \nn \\
&+&  \frac{q^2 y}{6 (1-4 y)^2 x_Z^3}  \bigg[ x_Z \left((4 y-1) x_Z \left(-4 y \left(8 a^2+4 a v+11 v^2\right)+2 a^2 (4
   y-1) x_Z+17 (a+v)^2\right) \right. \nn \\
&+& \left.   6 y \left(4 y \left(5 a^2+8 a v+7 v^2\right)-7
   (a+v)^2\right)\right)-24 y^2 (a+v)^2\bigg]   \mathcal B_0 \left( q^2, m^2, m^2 \right)  \nn \\
&+&   \frac{2 q^2 y}{3 (1-4 y)^2 x_Z^3}   \bigg[ x_Z \left(x_Z \left(-2 y \left(15 a^2+20 a v+v^2\right)+a^2 (8 y+1)
   x_Z+64 a^2 y^2+2 (a+v)^2\right) \right. \nn \\
&+&  \left.   y \left(7 (a+v)^2-4 y \left(10 a^2+8 a v+13
   v^2\right)\right)\right)+6 y^2 (a+v)^2\bigg]   \mathcal B_0 \left( q^2, m_Z^2, m_Z^2 \right)  \nn \\
&+&  \frac{q^2 y}{(1-4 y)^2 x_Z^2}   \bigg[ 2 x_Z \left(-4 y^2 \left(a^2-4 a v-4 v^2\right)-y \left(a^2+4 a v+10
   v^2\right)-4 a^2 y x_Z+(a+v)^2\right) \nn \\
&+&y \left(4 y \left(3 a^2-4 a v+3 v^2\right)+a^2+6 a
   v+v^2\right)\bigg]    \mathcal B_0 \left( m^2, m^2, m_Z^2 \right)  \nn \\
&+&  \frac{4 q^4 y^2}{(1-4 y)^2 x_Z^4}   \bigg[ x_Z \left(x_Z \left(x_Z \left(a^2 x_Z-a^2-2 v y (4 a+v)+v^2\right)-y
   \left(4 a^2 (4 y-1)-6 a v  \right.\right.\right.   \nn \\ 
&+& \left.\left.\left.     v^2 (8 y+1)\right)\right)+y \left(2 y \left(4 a^2+4 a v+5
   v^2\right)-(a+v)^2\right)\right)-y^2 (a+v)^2\bigg]       \mathcal C_0 \left( m^2, m_Z^2, m_Z^2 \right)   \nn \\
&+&  \frac{q^4 y}{(1-4 y)^2 x_Z^4}   \bigg[ x_Z \left((4 y-1) x_Z \left((4 y-1) x_Z \left(2 y
   \left(a^2+v^2\right)-(a+v)^2\right)-2 y^2 \left(7 a^2+8 a v  10 v^2\right)  \right.\right.   \nn \\
&+&  \left.\left.   6 y
   (a+v)^2\right)+y^2 \left(4 y \left(7 a^2+12 a v+9 v^2\right)-9 (a+v)^2\right)\right)-4 y^3
   (a+v)^2\bigg]      \mathcal C_0 \left( m_Z^2, m^2, m^2 \right) \,, \nn
\eea

\bea
f^Z_4(q^2) &=&
-\frac{8 y}{3 x_Z^2 \left(1 -4 y \right)^2} \bigg[ a^2 x_Z \left((2 y-3) x_Z-14 y+6\right)-5 c^2 y \left(x_Z-1\right)\bigg] \nn \\
&+&  \frac{4 (2 y-3) \left(2 a^2 x_Z+c^2\right)}{3 q^2 (1-4 y)^2 x_Z} \bigg[ \mathcal A_0 \left( m^2 \right) - \mathcal A_0 \left( m_Z^2 \right)  \bigg] 
 + \frac{4 y}{3 (4  y-1)^3 x_Z^3}   \bigg[ x_Z \left((4 y-1) x_Z   \right.  \nn \\
&\times& \left.  \left(a^2 (1-4 y) x_Z+6 a^2 (8 y-3)+c^2 (4  y-1)\right)+3 y \left(4 a^2 (3-7 y)+7 c^2 (1-4 y)\right)\right)+30 c^2 y^2 \bigg]    \nn \\
&\times&    \mathcal B_0( q^2, m^2, m^2) 
+  \frac{4 y}{3 (4 y-1)^3 x_Z^3}   \bigg[ x_Z \left(y \left(12 a^2 (7 y-3)+c^2 (68 y+13)\right)-x_Z \left(a^2 (16 y+11) x_Z  \right. \right.  \nn \\
&+& \left.\left.  2 a^2 \left(64 y^2-34 y-3\right)+c^2 (26 y+1)\right)\right)-30 c^2 y^2 \bigg]   \mathcal B_0 \left( q^2, m_Z^2, m_Z^2 \right)   \nn \\
&+&  \frac{4 y}{(4 y-1)^3 x_Z^2}   \bigg[ 2 a^2 x_Z \left((6 y+1) x_Z-8 y^2+4 y-3\right)-c^2 (8 (y-2) y+1)  \left(x_Z-1\right)\bigg]  \nn \\
&\times& \mathcal B_0( m^2, m^2, m_Z^2) 
- \frac{4 q^2 y}{(4 y-1)^3 x_Z^4}   \bigg[ a^2 x_Z \left((6 y+1) x_Z^3+6 y (2 y-3) x_Z^2+2 y ((17-40 y) y+2) x_Z  \right.   \nn \\
&+& \left.  4  y^2 (7 y-3)\right)-c^2 y \left(x_Z \left(x_Z \left(-4 (y+1) x_Z+2 y (8 y+11)+1\right)-6 y  (6 y+1)\right)+10 y^2\right) \bigg]   \nn \\
&\times&   \mathcal C_0 \left( m^2, m_Z^2, m_Z^2 \right) 
+   \frac{4 q^2 y}{(4 y-1)^3 x_Z^4}   \bigg[ x_Z \left((4 y-1) x_Z \left(3 y \left(2 a^2 (5 y-2)+c^2 (4 y-1)\right)  \right.\right. \nn \\
&-& \left. \left.  2  a^2 (y (12 y-7)+1) x_Z\right)+4 y^2 \left(a^2 (3-7 y)+3 c^2 (1-4 y)\right)\right)+10 c^2  y^3\bigg]     \mathcal C_0 \left( m_Z^2, m^2, m^2 \right) \,, \nn
\eea

\bea
f^Z_5(q^2) &=&
\frac{2 y \left(2 a^2 x_Z \left((8 y-3) x_Z-44 y+12\right)+c^2 \left((32 y-9) x_Z+4  y\right)\right)}{3 (1-4 y) x_Z^2} \nn \\
&+&  \frac{2 (8 y-3) \left(2 a^2 x_Z+c^2\right)}{3 q^2 (1- 4 y) x_Z} \bigg[ \mathcal A_0 \left( m_Z^2 \right) - \mathcal A_0 \left( m^2 \right) \bigg] 
+  \frac{2 y}{3 (1-4 y)^2 x_Z^3}  \bigg[x_Z \left((4 y-1) x_Z  \right. \nn \\
&\times&  \left.  \left(4 a^2 (1-4 y) x_Z+12 a^2 (3 y-1)+5 c^2 (1-4 y)\right)+24 a^2 (1-3 y) y\right)+12 c^2 y^2\bigg]  \mathcal B_0( q^2, m^2, m^2) \nn \\
&+&  \frac{4 y}{3 (1-4 y)^2 x_Z^3}  \bigg[ a^2 x_Z \left(x_Z \left((16 y-7) x_Z+4 y (8 y-5)+6\right)+12 y (3 y-1)\right)+c^2 \left((4 y+5) y x_Z \right. \nn \\
&+&   \left. (1-10 y) x_Z^2-6 y^2\right)\bigg]   \mathcal B_0 \left( q^2, m_Z^2, m_Z^2 \right) 
+  \frac{2 y}{\left(x_Z-4 y x_Z\right)^2}   \bigg[ 2 a^2 x_Z \left((2-4 y) x_Z+2 (7-12 y) y  \right. \nn \\
&-& \left.  3\right)+c^2 \left(2 y \left(y \left(8 x_Z+4\right)-5\right)+1\right)\bigg]   \mathcal B_0 \left( m^2, m^2, m_Z^2 \right) 
- \frac{4 q^2 y}{(1-4 y)^2 x_Z^4}   \bigg[ a^2 x_Z \left(x_Z \left((2 y-1) x_Z  \right. \right.  \nn \\ 
&\times&  \left.\left.  \left(6 y-x_Z\right)+6 (3-8 y)  y^2\right)+4 (3 y-1) y^2\right)+c^2 y \left(x_Z \left(x_Z \left(4 y x_Z+2 y (8 y-7)+1\right) \right. \right. \nn \\
&+& \left. \left.  2 y (2 y+1)\right)-2 y^2\right)\bigg]    \mathcal C_0 \left( m^2, m_Z^2, m_Z^2 \right) \nn \\
&+& \frac{4 q^2 y^2}{(1-4 y)^2 x_Z^4}  \left((4 y-1) x_Z-y\right) \bigg[Êx_Z \left(4 a^2 (3 y-1)+c^2 (1-4 y)\right)-2 c^2 y \bigg]    \mathcal C_0 \left( m_Z^2, m^2, m^2 \right) \nn \\
&+&  \chi \, \frac{32 a^2 y}{1 - 4 y} \bigg\{  \mathcal B_0 \left( q^2, m_Z^2, m_Z^2 \right) -  \mathcal B_0 \left( m^2, m^2, m_Z^2 \right) - \frac{q^2 (2 y - x_Z)}{2x_Z} \mathcal C_0 \left( m^2, m_Z^2, m_Z^2 \right)
\bigg\} \,, \nn 
\eea

\bea
f^Z_6(q^2) &=&
\frac{2\, a \, v \, y}{3 (4 y-1) x_Z^2}  \left((8 y-9) x_Z-8 y\right)
+ \frac{2 \, a \, v (8 y- 3)}{3 q^2 (1-4 y) x_Z} \bigg[ \mathcal A_0 \left( m^2 \right) - \mathcal A_0 \left( m_Z^2 \right) \bigg] \nn \\
&+&  \frac{2 \, a \, v \, y}{3 (1-4 y)^2 x_Z^3}  \bigg[ 6 (7-16 y) y x_Z+(4 y-1) (8 y-17) x_Z^2+24 y^2\bigg]   \mathcal B_0 \left( q^2, m^2, m^2 \right) \nn \\
&+&  \frac{8 \, a \, v \, y}{3 (1-4 y)^2 x_Z^3}  \bigg[ (16 y-7) y x_Z+(20 y-2) x_Z^2-6 y^2\bigg]    \mathcal B_0 \left( q^2, m_Z^2, m_Z^2 \right) \nn \\
&-& \frac{2 \, a \, v \, y}{(1-4 y)^2 x_Z^2}  \bigg[ (8 y+2) x_Z-2 y+1\bigg]  \mathcal B_0 \left( m^2, m^2, m_Z^2 \right) \nn \\
&+&  \frac{16 \, a \, v \, q^2 \, y^3}{(1-4 y)^2 x_Z^4}  \bigg[ x_Z \left(x_Z \left(4 x_Z-3\right)-4 y+1\right)+y\bigg]    \mathcal C_0 \left( m^2, m_Z^2, m_Z^2 \right) \nn \\
&+&  \frac{4 \, a \, v \, q^2 \, y}{(1-4 y)^2 x_Z^4}  \left( (4 y-1) x_Z-y\right) \bigg[ (8 y-5) y x_Z+(4 y-1) x_Z^2-4 y^2\bigg]  \mathcal C_0( m_Z^2, m^2, m^2) \,.
\eea
\normalsize

\subsection{The $W$ gauge boson contribution}
Finally we collect here the results for the $\hat T^{\mu\nu}_W$ matrix element. They are expressed in terms of scalar integrals and of the kinematic invariants $q^2$, $y = m^2/q^2$, $x_W = m^2/m_W^2$ and $x_f = m^2/m_f^2$, where $m_f$ is the mass of the fermion of flavor $f$ running in the loop. \\
As in the $Z$ boson case the conservation equation for the EMT implies the following relations among the form factors
\bea
\label{WIW1}
f^W_2 &=&  f^W_1 + q^2 f^W_6 + \frac{1}{4} \bigg[ \Sigma^R_W(m^2) -  \Sigma^L_W(m^2)  \bigg]  \,, \\
\label{WIW2}
f^W_3 &=& - q^2 f^W_5  - \frac{1}{2} \bigg[ \Sigma^L_W(m^2) + \Sigma^R_W(m^2) + 2 \Sigma^S_W(m^2) \bigg] \,,
\eea
which we have tested explicitly. Also in this case, as for the form factors with the exchange of a $Z$ boson, $f_5^W$, and hence 
$f_3^W$, depends on the parameter $\chi$.

We recall that the $O^{\mu\nu}_{C  3-6}$ operators are characterized by a linear mass suppression in the limit of small external fermion masses, while $O^{\mu\nu}_{C  2}$, even if not explicitly shown, has a quadratic suppression, which is present only in the $W$ case. Therefore, the leading contribution, in the limit of external massless fermions, is then given by the first form factor $f^W_1$ alone. \\

We present the explicit results for the $F^W_1$ and $F^W_4$ to $F^W_6$ in Appendix \ref{FWappendix}, while $F^W_2$ and $F^W_3$ can be computed using the Ward identities of Eq.(\ref{WIW1}) and (\ref{WIW2}). The form factors $f^W_k$ are obtained from $F^W_k$ multiplying by the CKM matrix elements and then summing over the fermion flavors, as explained in Eq.(\ref{Ftof}).

\section{Infrared singularities and soft bremsstrahlung}
\label{Sec.Infrared}
Here we provide a simple prove of the infrared safety 
of the $Tf\bar{f}$ vertex against soft radiative corrections and emissions
of massless gauge bosons.

An infrared divergence comes from the topology diagram depicted in Fig.\ref{diagrams}(a) with a virtual massless gauge boson exchanged between the two fermion lines and it is contained in the three-point scalar integral $\mathcal C_0(0,m^2,m^2)$. If we regularize the infrared singularity with a small photon (or gluon) mass $\lambda$ the divergent part of the scalar integral becomes
\bea
\mathcal C_0 (0, m^2, m^2) = \frac{x_s}{m^2 (1 - x_s^2)} \bigg\{  - 2 \log \frac{\lambda}{m}  \log x_s + \ldots \bigg\}
\eea 
where the dots stand for the finite terms not 
proportional to $\log \frac{\lambda}{m}$, and $x_s = - \frac{1-\beta}{1+\beta}$ with $\beta = \sqrt{1 - 4 m^2 / q^2}$ \,. \\
In the photon case the infrared singular part of the matrix element is then given by
\bea
\hat T^{\mu\nu}_{\gamma} &=& i \frac{\alpha}{4 \pi} Q^2 \bigg[ - \frac{1}{y}(2 y -1) \frac{x_s}{1-x_s^2} \log \frac{\lambda}{m}  \log x_s \bigg] \, \bar u(p_2) \, O^{\mu\nu}_{V  1} u(p_1)  + \ldots  \nn \\
&=& \frac{\alpha}{4 \pi} Q^2 \bigg[ - \frac{4}{y}(2 y -1) \frac{x_s}{1-x_s^2} \log \frac{\lambda}{m}  \log x_s \bigg] \, \hat T^{\mu\nu}_0 + \ldots \,,
\eea
which is manifestly proportional to the tree level vertex. 
On the other hand, the gluon contribution is easily obtained from the previous equation by replacing $\alpha \, Q^2$ with $\alpha_s \, C_2(N)$.

For the massless gauge boson contributions there is another infrared divergence coming from the renormalization counterterm. Its origin is in the field renormalization constants of charged particles arising from photonic or gluonic corrections to the fermion self energies. For example, in the photon case we have
\bea
\delta Z_L \bigg|_{\gamma}^{IR} = \delta Z_R \bigg|_{\gamma}^{IR} = - \frac{\alpha}{4 \pi} Q^2 \bigg\{ 4 \log \frac{\lambda}{m}  +  \ldots \bigg\} \,.
\eea 

However the processes described by the $\hat T^{\mu\nu}$ matrix element alone are not of direct physical relevance, since they cannot be distinguished experimentally from those involving the emission of soft massless gauge bosons. Adding incoherently the cross sections of all the different processes with arbitrary numbers of emitted soft photons (or gluons) all the infrared divergences are expected to cancel, as in an ordinary gauge theory \cite{Bloch:1937pw,Kinoshita:1962ur,
Lee:1964is}. This cancellation takes place between the virtual and the real bremsstrahlung corrections, and is valid to all orders in perturbation theory. In our case one has to consider only radiation of a single massless gauge boson with energy $k_0 < \Delta E$, smaller than a given cutoff parameter.
\begin{figure}[t]
\centering
\subfigure[]{\includegraphics[scale=0.8]{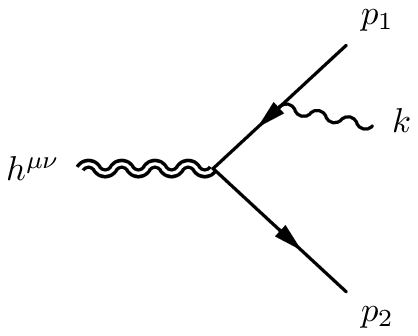}}  \hspace{2.5cm}
\subfigure[]{\includegraphics[scale=0.8]{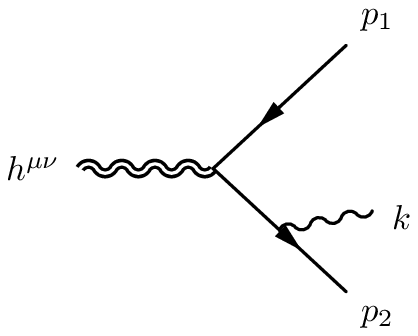}} 
\caption{Real emission diagrams of a massless gauge boson with momentum $k$. \label{realdiagrams}}
\end{figure}

For definiteness we consider the emission of a photon from the two external fermion legs. The gluon case, as already mentioned, is easily obtained from the final result with the replacement  $\alpha \, Q^2 \rightarrow \alpha_s \, C_2(N)$. In the soft photon approximation the real emission matrix element, corresponding to the sum of the two diagrams depicted in Fig. \ref{realdiagrams} is given by
\bea
\mathcal M_{soft} =   \mathcal M_0 \,  (e \, Q) \bigg[ \frac{\epsilon(k) \cdot p_1}{k \cdot p_1}  - \frac{\epsilon(k) \cdot p_2}{k \cdot p_2}  \bigg] \,,
\eea
where $k$ and $\epsilon(k)$ are the photon momentum and polarization vector respectively, while the sign difference between the two eikonal factors in the square brackets is due to the different fermion charge flow  of the diagrams. Here $\mathcal M_0$ is the Born amplitude which factorizes, in our case, as
\bea
\mathcal M_0 = A_{\mu\nu} \, \hat T_0^{\mu\nu} \,,
\eea
where $\hat T_0^{\mu\nu}$ is the tree level graviton vertex defined in Eq.(\ref{treeT}) and $A_{\mu\nu}$ is the remaining amplitude which does not participate to the soft photon emissions.

The cancellation of the infrared singularities occurs at the cross section level, therefore we have to square the soft photon matrix element, sum over the photon polarization and integrate over the soft photon phase space
\bea
d \sigma_{soft} = - d \sigma_0 \frac{\alpha}{2 \pi^2} Q^2 \int_{|\vec k | \le \Delta E} \frac{d^2 k}{2 k_0} \bigg[ \frac{p_1^2}{( k \cdot p_1)^2}  + \frac{p_1^2}{( k \cdot p_1)^2}  - 2 \frac{p_1 \cdot p_2}{k \cdot p_1 \, k \cdot p_2} \bigg]
\eea
where the infrared divergence is regularized by the photon mass $\lambda$ which appears through the photon energy $k_0 = \sqrt{|\vec k | + \lambda^2}$. \\
The generic soft integral
\bea
I_{i j} = \int_{|\vec k | \le \Delta E} \frac{d^2 k}{2 k_0}  \frac{2 p_i \cdot p_j}{k \cdot p_i \, k \cdot p_j} 
\eea
has been worked out explicitly in \cite{'tHooft:1978xw}, here we give only the infrared divergent parts needed in our case
\bea
I_{1 1} &=& I_{2 2} = 4 \pi \log \frac{\Delta E}{\lambda} + \ldots \,, \nn \\
I_{1 2} &=& - 8 \pi \left(1 - \frac{q^2}{2 m^2} \right) \frac{x_s}{1 - x_s^2}  \log \frac{\Delta E}{\lambda}  \log x_s + \ldots \,,
\eea
with $x_s = - \frac{1 - \beta}{1 + \beta}$ and $\beta = \sqrt{1 - 4m^2/q^2}$. \\
Using the previous results we obtain the infrared singular part of the soft cross section
\bea
d \sigma_{soft} = - d \sigma_0 \frac{\alpha}{4 \pi} Q^2 \bigg\{  8 \log \frac{\Delta E}{\lambda}  +  \frac{8}{y} (2 y -1)  \frac{x_s}{1-x_s^2}  \log \frac{\Delta E}{\lambda} \log x_s  + \ldots \bigg\} \,,
\eea
where dots stand for finite terms.

Exploiting the fact that the infrared divergences in the one-loop corrections and in the counterterm diagram multiply the tree level graviton vertex $\hat T^{\mu\nu}_0$, so that they contribute only with a term proportional to the Born cross section, we obtain
\bea
d \sigma_{virt} + d \sigma_{CT} + d \sigma_{soft} =   d \sigma_0  \, \frac{\alpha}{4 \pi} Q^2 \bigg[ 1  + \frac{1}{y} (2y-1) \frac{x_s}{1-x_s^2} \log x_s\bigg]  8 \log \frac{m}{\Delta E} + \ldots \,, 
\eea
for the photon case and an analog result for the gluon contribution. The sum of the renormalized virtual corrections with the real emission contributions is then finite in the limit $\lambda \rightarrow 0$.

\section{Conclusions} 
\label{Sec.Conclusions} 
We have computed  the one-loop EW and 
strong corrections to the 
flavor diagonal graviton-fermion vertices in the Standard Model.
This work is an extension, to the flavor diagonal case, 
of previous related study in which 
only the flavor-changing fermion graviton interactions had been investigated.
The result of our computation has been expressed in terms of 
a certain numbers of on-shell form factors, 
which have been given at leading order in the EW expansion and 
by retaining the exact dependence on the fermion masses.
We have also  included in our analysis the contribution of a 
non-minimally coupled Higgs sector, 
with an arbitrary value of the coupling parameter.
All these results can be easily extended to theories with fermion 
couplings to massive graviton, graviscalar and dilaton fields.

Moreover, we proved the infrared safety of the fermion-graviton vertices against radiative corrections of soft photons and gluons, where 
the ordinary cancellation mechanism
between the virtual and real bremsstrahlung corrections have been generalized
to the fermion-graviton interactions.

There are several phenomenological implications of this study that one could consider. Beside the possible applications to models with a low gravity scale, which would make the corrections discussed here far more sizeable, 
one could consider, for instance,  the specialization of our results to the neutrino sector, a definitely appealing argument on the cosmological side. Another possible extension would be to include,
 as a gravitational background, also a dilaton field, generated, for instance, from metric compactifications. We plan to return to some of these open issues in the future. 


\vspace{1cm}
\vbox{
\noindent{{\bf Acknowledgements} } \\
\noindent
E.G. and L.T. would like to thank the PH-TH division of CERN for its kind 
hospitality during the preparation of this work.
This work was supported by the ESF grant MTT60, 
by the recurrent financing SF0690030s09 project and by 
the European Union through the European Regional Development Fund.
}

\newpage 

\appendix
\section{Appendix. The structure of the W-boson form factors}
\label{FWappendix}

We collect in this appendix the expression of the form factors generated by the exchange of a W-boson in the loop. They are given by
\small
\bea
F^W_1(q^2) &=&
\frac{q^2 y \left(2 x_f^2 \left(3 x_W \left(y \left(x_W-10\right)+4\right)+8 y\right)+x_f x_W \left((6 y-3) x_W-8 y\right)-8 y \, x_W^2\right)}{6 (4 y-1) x_f^2 x_W^2}   \nn \\
&+&  \frac{y \left(x_f \left(3 x_W+2\right)+x_W\right)}{3 (1-4 y) x_f x_W} \bigg[  \mathcal A_0 \left(m_W^2 \right) -  \mathcal A_0 \left(m_f^2 \right) \bigg] 
+  \frac{q^2 y}{6 (1-4 y)^2 x_f^3 x_W^3} \bigg[ 2 x_f^3 \left(x_W \left(x_W \right. \right. \nn \\
&\times& \left.\left. \left(9 (2 y-1) y x_W-78 y^2+68 y-17\right)+42  y (2 y-1)\right)-24 y^2\right) 
+ x_f^2 x_W \left(2 (23  \right. \nn \\
&-& \left. 14 y) y x_W+(4 (5-12 y) y+1) x_W^2+72 y^2\right)+2 y (26 y-5) x_f x_W^3-24 y^2 x_W^3 \bigg]  \nn \\
&\times&   \mathcal B_0 \left( q^2, m_f^2, m_f^2 \right)
+  \frac{q^2 y}{3 (1-4 y)^2 x_f^3 x_W^3}  \bigg[ x_f^2 x_W \left(16 (y-1) y x_W+(4 y (3 y-1)+1) x_W^2 \right. \nn \\
&-& \left.  36 y^2\right) 
+ 4 x_f^3  \left(2 (y-1) x_W-3 y\right) \left((6 y-1) x_W-2 y\right)+12 (1-2 y) y x_f x_W^3+12 y^2  x_W^3 \bigg]   \nn \\ 
&\times& \mathcal B_0 \left( q^2, m_W^2, m_W^2 \right) 
+  \frac{q^2 y}{(1-4 y)^2 x_f^2  x_W^2}   \bigg[ x_f x_W \left((y (30 y-13)+4) x_f-2 y (y+1)\right) \nn \\
&-&  y x_W^2 \left(x_f \left((6 y-3) x_f-8 y+5\right) +2 (y+1)\right)+4 y (y+1) x_f^2 \bigg]     \mathcal B_0 \left( m^2, m_f^2, m_W^2 \right) \nn \\
&+&  \frac{2 q^4 y^2}{(1-4 y)^2 x_f^4 x_W^4} \bigg[  2 y x_f^3 x_W \left((8 y-2) x_f+5 y\right)+y x_f^2 x_W^2 \left(x_f \left(-4 (y-1) x_f-14 y+5\right) \right. \nn \\
&-&   \left. 6 y\right) 
 -2 y x_f x_W^3 \left(x_f \left(x_f \left(4 y x_f-3 y+3\right)-2 y+2\right)+y\right)-x_W^4 \left((2 y-1) x_f-2 y\right) \nn \\
&\times& \left(x_f \left(y x_f-2 y+1\right)+y\right)-4 y^2 x_f^4 \bigg]   \mathcal C_0 \left( m_f^2, m_W^2, m_W^2 \right) 
+ \frac{q^4 y}{(1-4 y)^2 x_f^4 x_W^4}   \bigg[ 2 y^2 x_f^3 x_W  \nn \\
&\times&   \left(9 (2 y-1) x_f+10 y\right)-y x_f^2 x_W^2 \left((34 y-25) y x_f+6 (y (9 y-8)+2) x_f^2+12 y^2\right) \nn \\
&-& y \left(x_f-1\right)^2 x_W^4  \left(\left(6 y^2-4 y+1\right) x_f^2+(2 y-1) y x_f-4 y^2\right)+2 x_f x_W^3 \left(4 (y-1) y^2 x_f  \right. \nn \\
&+& \left.  (2 y-1) \left(8 y^2-4 y+1\right) x_f^3+(3-2 y (y+2)) y x_f^2-2 y^3\right)-8 y^3 x_f^4 \bigg]   \mathcal C_0( m_W^2, m_f^2, m_f^2) \,, \nn
\eea

\bea
F^W_4(q^2) &=&
\frac{4 y}{3 (1-4 y)^2 x_f^2 x_W^2} \bigg[ x_f^2 \left((19 y-6) x_W-10 y\right)+x_f x_W \left((3-7 y) x_W+5 y\right)+5 y x_W^2\bigg] \nn \\
&+&  \frac{2 (2 y-3) \left(x_f \left(x_W+2\right)+x_W\right)}{3 q^2  (1-4  y)^2 x_f x_W} \bigg[ \mathcal A_0 \left( m_f^2 \right) - \mathcal A_0 \left( m_W^2 \right) \bigg] 
+  \frac{2 y}{3 (4 y-1)^3 x_f^3 x_W^3}  \bigg[ x_f^3 \left(x_W \right.  \nn \\
&\times& \left. \left(x_W \left(((5-6 y) y+1) x_W+y (108 y-77)+20\right)+6  (13-27 y) y\right)+60 y^2\right)+x_f^2 x_W \nn \\
&\times& \left((116 y-59) y x_W+2 (y (5 y-1)-1) x_W^2-90  y^2\right)+5 (1-10 y) y x_f x_W^3+30 y^2 x_W^3\bigg]    \nn \\
&\times&  \mathcal B_0 \left( q^2, m_f^2, m_f^2 \right)  
+  \frac{2 y}{3 (4 y-1)^3 x_f^3 x_W^3}  \bigg[ (x_W^3 \left(x_f \left(x_f \left((y-1) (6 y+1) x_f+14 (2   \right.\right.\right. \nn \\
&-& \left.\left.\left. 3 y) y  - 10\right)+3 y  (22 y-13)\right)-30 y^2\right)-(4 y-1) x_f^2 x_W^2 \left((7 y+4) x_f+25 y\right)   \nn \\
&+& 10 y x_f^2 x_W  \left((13 y-1) x_f+9 y\right)-60 y^2 x_f^3\bigg]   \mathcal B_0 \left( q^2, m_W^2, m_W^2 \right)  
+  \frac{2 y}{(4 y-1)^3 x_f^2 x_W^2}  \bigg[ (x_W^2  \nn \\
&\times& \left(\left(8 y^2-4 y  +  3\right) x_f-8 (y-2) y-1\right)-x_f x_W \left((24 (y-1) y+7) x_f+8 (y-2) y  \right.  \nn \\
&+&  \left. 1\right)+2 (8 (y-2) y   +  1) x_f^2\bigg]    \mathcal B_0 \left( m^2, m_f^2, m_W^2 \right)  
+   \frac{2 q^2 y}{(4 y-1)^3 x_f^4 x_W^4} \bigg[ -50 y^3 \left(x_f+1\right) \nn \\
&\times&   x_f^3 x_W+2 y x_f^2 x_W^2 \left(30 y^2 x_f   +  (y (11 y+5)-1) x_f^2+15 y^2\right)+y x_f x_W^3 \left(x_f \left(x_f \left((2 y (5 y \right.\right.\right.  \nn \\
&-&  \left.\left.\left.  6)+3) x_f+22 y^2-26 y+7\right)  +   6 (3-7 y) y\right)+10 y^2\right)-x_W^4 \left(4 (2-3 y) y x_f+(2 (y  \right. \nn \\ 
&-&  \left. 1) y+1) x_f^2+10 y^2\right) \left(x_f \left(y x_f-2 y   +  1\right)+y\right)+20 y^3  x_f^4\bigg]    \mathcal C_0 \left( m_f^2, m_W^2, m_W^2 \right)  \nn \\
&+&  \frac{2 q^2 y}{(4 y-1)^3 x_f^4 x_W^4}   \bigg[-2 y^2 x_f^3 x_W \left((37 y-18) x_f  
+  25 y\right)+6 y x_f^2 x_W^2  \left((16 y-9) y x_f   \right.  \nn \\
&+& \left.   3 (y (5 y-4)+1) x_f^2+5 y^2\right)+y \left(x_f-1\right)^2 x_W^4  \left((2 (y-1) y+1) x_f^2-10 y^2\right) \nn \\
&+&  x_f x_W^3 \left(6 (3-7 y) y^2 x_f+\left(-26 y^2+28  y-11\right) y x_f^2+\left(y \left(-38 y^2+34 y-11\right)+2\right) x_f^3 \right.  \nn \\
&+& \left.   10 y^3\right)+20  y^3 x_f^4\bigg]    \mathcal C_0 \left( m_W^2, m_f^2, m_f^2 \right)  \,, \nn
\eea

\bea
F^W_5(q^2) &=&
\frac{y}{3 (4 y-1) x_f^2 x_W^2}   \bigg[ x_f^2 \left(x_W+2\right) \left(3 (4 y-1) x_W-4 y\right)+x_f x_W \left((9-32 y) x_W+4 y\right) \nn \\
&+&   4 y x_W^2\bigg] 
+  \frac{(8 y-3) \left(x_f \left(x_W+2\right)+x_W\right)}{3 q^2 (1-4 y) x_f x_W} \bigg[  \mathcal A_0 \left( m_W^2 \right)  -  \mathcal A_0 \left( m_f^2 \right) \bigg] \nn \\
&+&  \frac{y}{3 (1-4 y)^2 x_f^3 x_W^3}    \bigg[x_f^3 \left(x_W+2\right) \left(12 y^2 \left(x_W-1\right)^2-4 y  \left(x_W-3\right) x_W+x_W^2\right)-x_f^2 x_W \nn \\
&\times&   \left(8 (2 y+1) y x_W+(4 y (11 y-7)+5)  x_W^2+36 y^2\right)+4 (2-11 y) y x_f x_W^3+12 y^2 x_W^3\bigg]   \nn \\
&\times&  \mathcal B_0( q^2, m_f^2, m_f^2) 
+  \frac{2 y}{3 (1-4 y)^2 x_f^3 x_W^3}   \bigg[  x_W^3 \left(x_f \left(x_f \left((y (6 y+5)-2) x_f+2 (10-9 y) y \right.\right.\right.  \nn \\
&-& \left.\left.\left.   5\right)+9 y  (2 y-1)\right)-6 y^2\right)+(4 y-1) x_f^2 x_W^2 \left((7 y-8) x_f+y\right)+2 y x_f^2 x_W  \left((13 y \right. \nn \\
&-& \left.  1) x_f+9 y\right)-12 y^2 x_f^3 \bigg]    \mathcal B_0 \left( q^2, m_W^2, m_W^2 \right)  
+  \frac{y}{(1-4 y)^2 x_f^2 x_W^2}    \bigg[ x_f^2 \left(x_W \left((1-2 y (4 y  \right.\right.  \nn \\
&+& \left.\left.   1)) x_W+2 (9-4 y) y-5\right)+4 y (4  y-5)+2\right)+x_f x_W \left(4 (1-2 y)^2 x_W+2 (5-4 y) y  \right.  \nn \\
&-&  \left. 1\right)+(2 (5-4 y) y-1) x_W^2\bigg]     \mathcal B_0( m^2, m_f^2, m_W^2)  
-  \frac{2 q^2 y}{(1-4 y)^2 x_f^4  x_W^4}   \bigg[ 10 y^3 \left(x_f+1\right) x_f^3 x_W  \nn \\
&-&    2 y x_f^2 x_W^2 \left((2 y+1) y   x_f+(y (3 y+4)-1) x_f^2+3 y^2\right)-y \left(x_f+1\right) x_f x_W^3 \left(2 (1 \right.    \nn \\
&-&  \left.     2 y) y  x_f+(2 (y-6) y+3) x_f^2+2 y^2\right)+x_W^4 \left(2 (1-2 y) y x_f+(2 (y-2) y+1) x_f^2   \right. \nn \\
&+&\left.    2  y^2\right) \left(x_f \left(y x_f-2 y+1\right)+y\right)-4 y^3 x_f^4\bigg]    \mathcal C_0( m_f^2, m_W^2, m_W^2) 
- \frac{2 q^2 y^2}{(1-4 y)^2  x_f^4 x_W^4}  \bigg[ 2 x_f x_W  \nn \\
&\times&   \left((y-1) x_f+y\right)+\left(x_f-1\right) x_W^2 \left((2  y-1) x_f-2 y\right)-4 y x_f^2\bigg] \bigg[ x_f^2 \left(y  \left(x_W-1\right)^2+x_W\right)   \nn \\
&-&   2 y x_f x_W \left(x_W+1\right)+y x_W^2\bigg]   \mathcal C_0 \left( m_W^2, m_f^2, m_f^2 \right) 
+  \chi \bigg\{   \frac{8 y \left(x_f+1\right)}{(1-4 y) x_f} \bigg[  \mathcal B_0 \left( q^2, m_W^2, m_W^2 \right)  \nn \\
&-&  \mathcal B_0 \left( m^2, m_f^2, m_W^2 \right)     \bigg] 
+  \frac{8 q^2 y}{(4 y-1) x_f^2 x_W}   \bigg[ y x_f \left(x_f+1\right)-x_W \left(x_f \left(y x_f-2   y+1\right)+y\right)\bigg]  \nn \\
&\times&  \mathcal C_0 \left( m_f^2, m_W^2, m_W^2 \right)  \bigg\} \,, \nn
\eea

\bea
F^W_6(q^2) &=&
\frac{y}{6 (1-4 y) x_f^2 x_W^2}  \bigg[-2 x_f x_W \left((4 y-9) x_f+4 y\right)+\left(x_f-1\right) x_W^2 \left(3 x_f+8
   y\right)+16 y x_f^2\bigg] \nn \\
&+&  \frac{(8 y-3) \left(x_f \left(x_W-2\right)-x_W\right)}{6 q^2 (1 - 4 y) x_f x_W} \bigg[ \mathcal A_0 \left( m_W^2 \right)  -  \mathcal A_0 \left( m_f^2 \right) \bigg] 
-   \frac{y}{6 (1-4 y)^2 x_f^3 x_W^3}    \bigg[\left(x_f-1\right) \nn \\
&\times&   x_W^3 \left(2 (5-8 y) y x_f+(2 y (12 y-7)-1) x_f^2+24  y^2\right)+2 x_f^2 x_W^2 \left(((53-48 y) y-17) x_f \right. \nn \\
&+& \left.  y (16 y+23)\right)+12 y x_f^2 x_W \left((10 y-7) x_f+6 y\right)-48 y^2 x_f^3\bigg]    \mathcal B_0 \left( q^2, m_f^2, m_f^2 \right)  \nn \\
&-&  \frac{y}{3 (1-4 y)^2 x_f^3 x_W^3}    \bigg[ -\left(x_f-1\right) x_W^3 \left(12 (1-2 y) y x_f+(4 y (3 y-1)+1) x_f^2+12 y^2\right) \nn \\
&+&   8 x_f^2 x_W^2 \left((4 (y-2) y+1) x_f-2 y (2 y+1)\right)-4 y x_f^2 x_W \left(7  (y-1) x_f+9 y\right)+24 y^2 x_f^3\bigg]    \nn \\
&\times&   \mathcal B_0 \left( q^2, m_W^2, m_W^2 \right) 
-  \frac{y}{2 (1-4 y)^2 x_f^2 x_W^2}   \bigg[ x_f^2 \left(x_W \left(2 y \left(x_W+7\right)+x_W+5\right)-4 y+2\right) \nn \\
&+&  x_f x_W  \left(-4 y x_W+2 y-1\right)+(2 y-1) x_W^2\bigg]     \mathcal B_0 \left( m^2, m_f^2, m_W^2 \right)  
-  \frac{2 q^2 y^2}{(1-4 y)^2 x_f^4 x_W^4}    \bigg[ 2 y x_f^3  \nn \\
&\times&    x_W \left((3 y-2) x_f+5 y\right)+y x_f^2 x_W^2 \left(x_f   \left((2 y+7) x_f+4 y+5\right)-6 y\right)-2 y x_f x_W^3 \left(x_f \left(x_f  \right.\right. \nn \\
&\times& \left.\left.   \left((3 y+1)  x_f-5 y+5\right)+y+2\right)+y\right)+\left(x_f-1\right) x_W^4 \left((2 y-1) x_f-2 y\right)  \left(x_f \left(y x_f   \right.\right.  \nn \\
&-& \left.\left.   2 y+1\right)+y\right)-4 y^2 x_f^4\bigg]   \mathcal C_0 \left( m_f^2, m_W^2, m_W^2 \right) 
-   \frac{q^2 y}{(1-4 y)^2 x_f^4 x_W^4}    \bigg[ 2 y^2 x_f^3 x_W \left((14 y-9) x_f \right. \nn \\
&+&  \left.  10 y\right)-y x_f^2 x_W^2 \left((16  y-25) y x_f+3 (y (12 y-13)+4) x_f^2+12 y^2\right)-y \left(x_f-1\right)^2 x_W^4 \nn \\
&\times&  \left(-y  x_f+(y (4 y-3)+1) x_f^2-4 y^2\right)+2 x_f x_W^3 \left(x_f \left(x_f \left((y (2 y (5  y-6)+5)-1) x_f \right.\right.\right. \nn \\
&-& \left.\left.\left.   6 y^3+3 y\right)-2 y^2 (y+2)\right)-2 y^3\right)-8 y^3 x_f^4\bigg]     \mathcal C_0 \left( m_W^2, m_f^2, m_f^2 \right) \,.
\eea
\normalsize

\section{Appendix. The Ward identity} 
\label{WIS}
In this appendix we fill-in some of the gaps in the derivation of the Ward identity satisfied by the effective action of the Standard Model.
In order to simplify our notations, as mentioned in the main section, we take as an illustration a theory with a spin-1 gauge field $(A_\mu)$, a charged scalar $(\phi)$ and a fermion $(\psi)$. In the case of the Standard Model the extension of this analysis is straightforward but quite lengthy. We have replicas of the spin-1's  (i.e. $A_\mu^a\equiv( A_g, A_\gamma, W^i, Z)$)  on which we implicitly sum over, while 
the scalar $\phi$ summarizes both the Higgs and the ghost $(\omega^{i})$ contributions $(\phi\equiv(\mathcal{H},\omega^{i})$, 
$(\phi^\dagger\equiv(\mathcal{H}^\dagger, \bar{\omega}^{i}))$ and $\psi$ all the fermions.  

In these condensed notations, the conservation equation of the EMT of the Standard Model takes the following off-shell form
\bea
\partial^{\mu}T_{\mu\nu} &=& -\frac{\delta S}{\delta \psi} \partial_{\nu}\psi - \partial_{\nu}\bar \psi \frac{\delta S}{\delta \bar \psi} + \frac{1}{2}\partial^{\mu}\left( \frac{\delta S}{\delta \psi}\sigma_{\mu\nu}\psi - \bar \psi \sigma_{\mu\nu}\frac{\delta S}{\delta \bar \psi} \right) - \partial_{\nu}A_{\mu}^a \frac{\delta S}{\delta A_{\mu}^a} \nn\\
&+& \partial_{\mu}\left( A_{\nu}^a \frac{\delta S}{\delta A_{\mu}^a} \right) - \frac{\delta S}{\delta \phi} \partial_{\nu} \phi - \partial_{\nu}\phi^\dagger \frac{\delta S}{\delta \phi^\dagger} \,  \label{EMTdivergence}
\eea
where we are implicitly summing on all the spin-1, scalars and fermions. 
The off-shell relation (\ref{EMTdivergence}) can be inserted into the functional integral in order to derive some of the Ward identities satisfied by the $T f\bar{f}$ correlator. Whence we define the generating functional of the theory in the fluctuations of the background gravitational metric $(h_{\mu\nu})$

\bea
&& Z[J,\eta,\bar\eta,\zeta,\zeta^\dag,h_{\mu\nu}] =
\int \mathcal D A \, \mathcal D \psi \, \mathcal D \bar\psi \,
\mathcal D \phi \, \mathcal D \phi^\dagger \, \exp\bigg\{ i \int d^4
x \left( \mathcal{L}_{SM} + J_{\mu}A^{\mu} \right. \nn \\
&& \hspace{6cm} \left.  + \bar \eta \psi +
\bar \psi \eta + \zeta^\dag \phi + \phi^\dagger \zeta +
h_{\mu\nu}T^{\mu\nu}\right) \bigg\} \,.
\eea
The EMT, obviously, is chosen to be the symmetric one and on-shell conserved. We have denoted with $J,\eta,\bar\eta,\zeta,\zeta^\dag$ the
sources of the gauge field(s), the fermion and antifermion fields and scalar and its conjugate respectively. The generating functional $W$ of the connected Green's functions is, as usual, denoted by 
\bea 
\exp i \, W[J,\eta,\bar\eta, \zeta, \zeta^\dag,h_{\mu\nu}] =
\frac{Z[J,\eta,\bar\eta, \zeta, \zeta^\dag,h_{\mu\nu}]}{Z[0]}
\eea
(normalized to the vacuum functional). The effective action, defined as the generating functional $\Gamma$ of the 1-particle irreducible amplitudes is obtained from $W$ by a Legendre transformation with respect to all the sources, except, in our case, the metric fluctuation $h_{\mu\nu}$, which is taken as a background external field
\bea
\Gamma[A_c,\bar \psi_c, \psi_c,
\phi_c^\dag, \phi_c, h_{\mu\nu}] = W[J,\eta,\bar\eta, \zeta, \zeta^\dag,h_{\mu\nu}] -
\int d^4 x \left( J_{\mu}A^{\mu}_c + \bar \eta \psi_c + \bar \psi_c \eta
+ \zeta^\dag \phi_c + \phi_c^\dag \zeta \right).
\label{1PIfunctional}
\eea
As usual, we eliminate the source fields  from the right hand side of Eq.~(\ref{1PIfunctional}) inverting the relations
\bea
A^{\mu}_c = \frac{\delta W }{\delta J_{\mu}}, \qquad
\psi_c = \frac{\delta W }{\delta \bar \eta}, \qquad \bar \psi_c =
\frac{\delta W }{\delta \eta}, \qquad \phi_c = \frac{\delta W
}{\delta  \zeta^\dag}, \qquad  \phi_c^\dag = \frac{\delta W }{\delta
\zeta} \label{Legendre1}
\eea
so that the functional derivatives of the effective action $\Gamma$ with respect to its independent variables are
\bea
\frac{\delta\Gamma}{\delta A^{\mu}_c} = - J_{\mu}, \qquad
\frac{\delta\Gamma}{\delta \psi_c} = - \bar \eta, \qquad
\frac{\delta\Gamma}{\delta \bar \psi_c} = - \eta, \qquad
\frac{\delta\Gamma}{\delta \phi_c} = - \zeta^\dagger, \qquad
\frac{\delta\Gamma}{\delta \phi_c^\dagger} = - \zeta,
 \label{Legendre2}
\eea
and for the source $h_{\mu\nu}$ we have instead
\bea
\frac{\delta\Gamma}{\delta h_{\mu\nu}} = \frac{\delta W}{\delta h_{\mu\nu}}. \label{Legendre3}
\eea
The conservation of the EMT given by Eq.~(\ref{EMTdivergence}) is rewritten in terms of classical fields and then re-expressed in functional form by differentiating $W$ with respect to $h_{\mu\nu}$. We use Eq.~(\ref{EMTdivergence})
under the functional integral. We obtain
\bea
\partial_{\mu} \frac{\delta W}{\delta h_{\mu\nu}} &=&  \bar \eta \, \partial_{\nu} \frac{\delta W}{\delta \bar \eta} +
\partial_{\nu} \frac{\delta W}{\delta \eta} \eta - \frac{1}{2}\partial^{\mu}\left(\bar \eta \sigma_{\mu\nu} \frac{\delta W}{\delta \bar \eta}-
 \frac{\delta W}{\delta \eta}\sigma_{\mu\nu}\eta  \right) + \partial_{\nu} \frac{\delta W}{\delta J_{\mu}} J_{\mu}
 - \partial_{\mu} \left( \frac{\delta W}{\delta J_{\mu} }J_{\nu} \right) \nn \\
 &+& \zeta^\dagger \partial_{\nu}\frac{\delta W}{\delta \zeta^\dagger} + \partial_{\nu}\frac{\delta W}{\delta \zeta} \zeta \,,
\label{divW}
\eea
and finally, for the one particle irreducible generating functional, this gives Eq. (\ref{classical}), after using 
Eq.~(\ref{divW}) with the help of Eqs.~(\ref{Legendre1}), (\ref{Legendre2}), (\ref{Legendre3}).

\section{Feynman rules}
\label{feynrules}

We collect here all the Feynman rules involving a graviton that have been used in this work. All the momenta are incoming

\begin{itemize}
\item{ graviton - gauge boson - gauge boson vertex}
\\ \\
\begin{minipage}{95pt}
\includegraphics[scale=1.0]{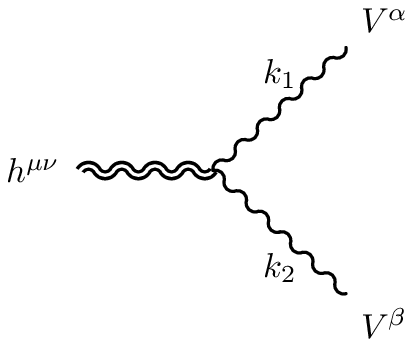}
\end{minipage}
\begin{minipage}{70pt}
\bea
= - i \frac{\kappa}{2} \bigg\{ \left( k_1 \cdot k_2  + M_V^2 \right) C^{\mu\nu\alpha\beta}
+ D^{\mu\nu\alpha\beta}(k_1,k_2) + \frac{1}{\xi}E^{\mu\nu\alpha\beta}(k_1,k_2) \bigg\}
\nn
\eea
\end{minipage}
\bea
\label{FRhVV}
\eea
where $V$ stands for the vector gauge bosons $g$, $\gamma$, $Z$ and $W$.
\item{graviton - fermion - fermion vertex}
\\ \\
\begin{minipage}{95pt}
\includegraphics[scale=1.0]{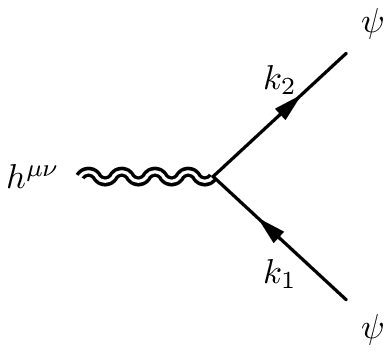}
\end{minipage}
\begin{minipage}{70pt}
\bea
=- i \frac{\kappa}{8} \bigg\{ \gamma^\mu \, (k_1 + k_2)^\nu + \gamma^\nu \,(k_1 + k_2)^\mu - 2 \, \eta^{\mu\nu} \left( \ksl_1 + \ksl_2 - 2 m_f \right)\bigg\}
\nn
\eea
\end{minipage}
\bea
\label{FRhFF}
\eea
\item{graviton - scalar - scalar vertex}
\\ \\
\begin{minipage}{95pt}
\includegraphics[scale=1.0]{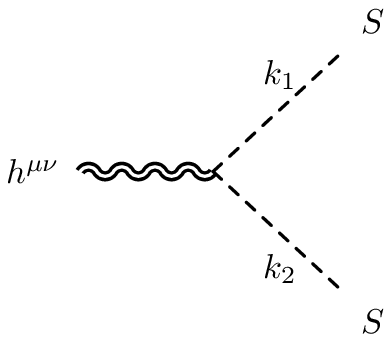}
\end{minipage}
\begin{minipage}{70pt}
\bea
&=&  i \frac{\kappa}{2} \bigg\{ k_{1\, \rho} \, k_{2 \, \sigma} \, C^{\mu\nu\rho\sigma}  - M_S^2 \, \eta^{\mu\nu} \bigg\} \nn \\
&-&  i \frac{\kappa}{2}  2 \chi  \bigg\{ (k_1+k_2)^{\mu}(k_1+k_2)^{\nu} - \eta^{\mu\nu} (k_1+k_2)^2 \bigg\} \nn
\eea
\end{minipage}
\bea
\label{FRhSS}
\eea
where $S$ stands for the Higgs $H$ and the Goldstones $\phi$ and  $\phi^{\pm}$. The first line is the contribution coming from the minimal energy-momentum tensor while the second is due to the improvement term.
\item{graviton - scalar - fermion - fermion vertex}
\\ \\
\begin{minipage}{95pt}
\includegraphics[scale=1.0]{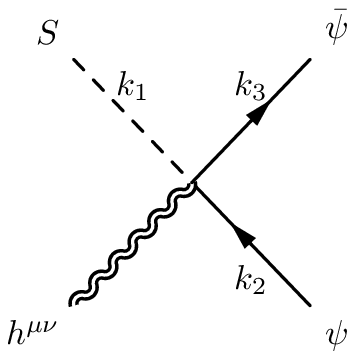}
\end{minipage}
\begin{minipage}{70pt}
\bea
=  \frac{\kappa}{2} \left( C^L_{S\bar \psi \psi} \, P_L + C^R_{S\bar \psi \psi} \, P_R \right) \, \eta^{\mu\nu}
\nn
\eea
\end{minipage}
\bea
\label{FRhSFF}
\eea
where  the coefficients are defined as
\bea
&& C^L_{h \bar \psi \psi}  = C^R_{h \bar \psi \psi} = - i \frac{e}{2 s_W} \frac{m}{m_W}  \,, \qquad
C^L_{\phi \bar \psi \psi}  = - C^R_{\phi \bar \psi \psi} =  i\frac{e}{2 s_W} \frac{m}{m_W} 2 I_3 \,, \nn \\
&& C^L_{\phi^+ \bar \psi \psi} = i \frac{e}{\sqrt{2} s_W} \frac{m_{\bar \psi}}{m_W} V_{\bar \psi \psi} \,, \qquad  C^R_{\phi^+ \bar \psi \psi} = - i \frac{e}{\sqrt{2} s_W} \frac{m_{\psi}}{m_W} V_{\bar \psi \psi} \,, \nn \\
&& C^L_{\phi^- \bar \psi \psi} = - i \frac{e}{\sqrt{2} s_W} \frac{m_{\bar \psi}}{m_W} V^*_{\bar \psi \psi} \,, \qquad  C^R_{\phi^- \bar \psi \psi} =  i \frac{e}{\sqrt{2} s_W} \frac{m_{\psi}}{m_W} V^*_{\bar \psi \psi} \,.
\eea
\item{graviton - gauge boson - fermion - fermion vertex}
\\ \\
\begin{minipage}{95pt}
\includegraphics[scale=1.0]{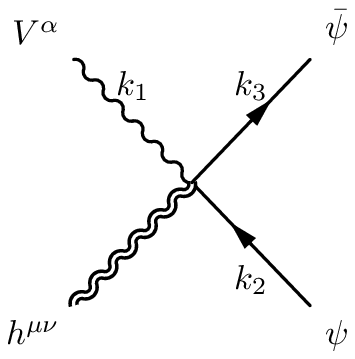}
\end{minipage}
\begin{minipage}{70pt}
\bea
= - \frac{\kappa}{2} \left( C^L_{V \bar\psi \psi} \, P_L + C^R_{V \bar\psi \psi} \, P_R\right) C^{\mu\nu\alpha\beta} \gamma_\beta
\nn
\eea
\end{minipage}
\bea
\label{FRhAFF}
\eea
with
\bea
&& C^L_{g \bar\psi \psi} = C^R_{g \bar\psi \psi} = i g_s T^a \,, \qquad  C^L_{\gamma \bar\psi \psi} = C^R_{\gamma \bar\psi \psi} = i e Q \,, \nn \\
&& C^L_{Z \bar\psi \psi} = i \frac{e}{2 s_W c_W} (v + a) \,, \quad C^R_{Z \bar\psi \psi} = i \frac{e}{2 s_W c_W} (v - a) \,, \nn \\
&& C^L_{W^+ \bar\psi \psi} = i \frac{e}{\sqrt{2} s_W} V_{\bar\psi \psi}  \,, \quad C^L_{W^- \bar\psi \psi} = i \frac{e}{\sqrt{2} s_W} V^*_{\bar\psi \psi}  \,, \quad C^R_{W^\pm \bar\psi \psi} = 0 \,,
\eea
and $v = I_3 - 2 s_W^2 Q$, $a = I_3$.
\end{itemize}
The tensor structures $C$, $D$ and $E$ which appear in the Feynman rules defined above are given by
\bea
&& C_{\mu\nu\rho\sigma} = \eta_{\mu\rho}\, \eta_{\nu\sigma} +\eta_{\mu\sigma} \, \eta_{\nu\rho} -\eta_{\mu\nu} \, \eta_{\rho\sigma} \,, \nn \\
&& D_{\mu\nu\rho\sigma} (k_1, k_2) = \eta_{\mu\nu} \, k_{1 \, \sigma}\, k_{2 \, \rho} - \biggl[\eta^{\mu\sigma} k_1^{\nu} k_2^{\rho} + \eta_{\mu\rho} \, k_{1 \, \sigma} \, k_{2 \, \nu}
  - \eta_{\rho\sigma} \, k_{1 \, \mu} \, k_{2 \, \nu}  + (\mu\leftrightarrow\nu)\biggr] \,, \nn \\
&& E_{\mu\nu\rho\sigma} (k_1, k_2) = \eta_{\mu\nu} \, (k_{1 \, \rho} \, k_{1 \, \sigma} +k_{2 \, \rho} \, k_{2 \, \sigma} +k_{1 \, \rho} \, k_{2 \, \sigma})
-\biggl[\eta_{\nu\sigma} \, k_{1 \, \mu} \, k_{1 \, \rho} +\eta_{\nu\rho} \, k_{2 \, \mu} \, k_{2 \, \sigma} +(\mu\leftrightarrow\nu)\biggr] \,. \nn \\
\eea

\section{Fermion self-energy}
\label{selfenergies}

The one-loop fermion two-point function, diagonal in the flavor space, is defined as
\bea
\Gamma_{\bar f f}(p) = i \bigg[ \psl P_L \, \Sigma^L(p^2) +  \psl P_R \, \Sigma^R(p^2)  + m \, \Sigma^S(p^2) \bigg]
\eea
where the three components $\Sigma^X(p^2)$, with $X = L,R,S$, are eventually given by the gluon, the photon, the Higgs, the Z and the W contributions
\bea
\Sigma^X(p^2) = \frac{\alpha_s}{4 \pi} C_2(N) \, \Sigma^X_g (p^2) + \frac{\alpha}{4 \pi} Q^2 \, \Sigma^X_{\gamma}(p^2) + \frac{G_F}{16 \pi^2 \sqrt{2}} \bigg[ m^2 \, \Sigma^X_h (p^2) +   \Sigma^X_Z (p^2) +  \Sigma^X_W(p^2) \bigg] \,.
\eea
The  $\Sigma^X(p^2)$ coefficients of the fermion self-energies are explicitly given by
\bea
&&  \Sigma^L_g (p^2)  = \Sigma^R_g (p^2) = \Sigma^L_\gamma (p^2)  = \Sigma^R_\gamma (p^2) =  - 2 \, \mathcal B_1 \left( p^2, m^2, 0 \right) - 1   \,, \nn   \\
&& \Sigma^S_g (p^2)  = \Sigma^S_\gamma (p^2) =  - 4 \, \mathcal B_0 \left( p^2, m^2, 0 \right) + 2     \,, \nn  \\
&& \Sigma^L_h (p^2) =  \Sigma^R_h (p^2) = - 2 \, \mathcal B_1 \left( p^2, m^2, m_h^2 \right)   \,, \nn \\
&& \Sigma^S_h (p^2) = 2 \, \mathcal B_0 \left( p^2, m^2, m_h^2 \right)   \,, \nn \\
&& \Sigma^L_W (p^2) = - 4 \sum_f V_{if}^* V_{f_i}  \bigg[ \left( m_f^2 + 2 m_W^2 \right) \mathcal B_1 \left( p^2, m_f^2, m_W^2 \right) + m_W^2 \bigg]   \,, \nn \\
&& \Sigma^R_W (p^2) = - 4 m^2 \sum_f V_{if}^* V_{f_i}  \,  \mathcal B_1 \left( p^2, m_f^2, m_W^2 \right)   \,, \nn  \\
&& \Sigma^S_W (p^2) = - 4  \sum_f V_{if}^* V_{f_i}  \, m_f^2 \, \mathcal B_0 \left( p^2, m_f^2, m_W^2 \right) \,, \nn \\
&& \Sigma^L_Z (p^2) = -2 m_Z^2 (v + a)^2 \bigg[ 2 \, \mathcal B_1 \left( p^2, m^2, m_Z^2 \right)  +1 \bigg]  - 2 m^2 \, \mathcal B_1 \left( p^2, m^2, m_Z^2 \right)  \,, \nn \\
&& \Sigma^R_Z (p^2) = -2 m_Z^2 (v - a)^2 \bigg[ 2 \, \mathcal B_1 \left( p^2, m^2, m_Z^2 \right)  +1 \bigg]  - 2 m^2 \, \mathcal B_1 \left( p^2, m^2, m_Z^2 \right)  \,, \nn \\
&& \Sigma^S_Z (p^2) = -2 m_Z^2 (v^2 -a^2) \bigg[ 4 \, \mathcal B_0 \left( p^2, m^2, m_Z^2 \right) -2 \bigg] - 2 m^2 \, \mathcal B_0 \left( p^2, m^2, m_Z^2 \right) \,,
\eea
where $v$ and $a$ are the vector and axial-vector $Z$-fermion couplings defined in Eq.(\ref{Zfermconst}) and
\bea
\mathcal B_1 \left( p^2, m_0^2, m_1^2 \right) = \frac{m_1^2 -m_0^2}{2 p^2} \bigg[ \mathcal B_0(p^2, m_0^2, m_1^2) -  \mathcal B_0(0, m_0^2, m_1^2) \bigg] -\frac{1}{2} \mathcal B_0(p^2, m_0^2, m_1^2) \,.
\eea

\section{Scalar integrals}
\label{scalarint}
In this Appendix we collect the definitions of the scalar integrals appearing in the computation of the matrix element. One-, two- and three- point functions are denoted respectively as $\mathcal A_0$, $\mathcal B_0$ and $\mathcal C_0$ with
\bea
\mathcal A_0(m_0^2) &=& \frac{1}{i \pi^2} \int d^n l \frac{1}{l^2 - m_0^2} \,, \nn \\
\mathcal B_0(p_1^2, m_0^2, m_1^2) &=& \frac{1}{i \pi^2} \int d^n l \frac{1}{(l^2 -m_0^2)((l+p_1)^2 -m_1^2)} \,, \nn \\ 
\mathcal C_0(p_1^2, (p_1-p_2)^2, p_2^2,m_0^2, m_1^2, m_2^2) &=&  \frac{1}{i \pi^2} \int d^n l \frac{1}{(l^2 -m_0^2)((l+p_1)^2 -m_1^2)((l+p_2)^2 -m_2^2)} \,. \nn \\
\eea
Because the kinematic invariants on the external states of our computation are fixed, $q^2 = (p_1 - p_2)^2$, $p_1^2 = p_2^2 = m^2$, we have defined the shorter notation for the three-point scalar integrals
\bea
\mathcal C_0(m_0^2, m_1^2, m_2^2) = \mathcal C_0(m^2, q^2, m^2, m_0^2, m_1^2, m_2^2) \,,
\eea
with the first three variables omitted.

\bibliographystyle{h-physrev5}

\end{document}